\documentclass[groupedaddress, showpacs, nofootinbib, amsmath,amssymb, aps, notitlepage]{revtex4-1}

\usepackage{subcaption}
\usepackage{float}
\usepackage{graphicx}
\usepackage{dcolumn}
\usepackage{bm}
\usepackage{hyperref}

\usepackage{xcolor}

\usepackage{multirow}

\allowdisplaybreaks[1]

\def\eq#1{{Eq.~(\ref{#1})}}
\def\fig#1{{Fig.~\ref{#1}}}
\newcommand{\ben}{\begin{eqnarray*}}
\newcommand{\een}{\end{eqnarray*}}
\newcommand{\un}[1]{\underline{#1}}

\newcommand{\pd}{\partial}
\newcommand{\ul}[1]{\underline{#1}}

\newcommand{\tr}{\mbox{tr}}

\newcommand{\as}{\alpha_s}

\begin{document}

\title{Helicity at Small $x$: Oscillations Generated by Bringing Back the Quarks}

\author{Yuri V. Kovchegov} 
         \email[Email: ]{kovchegov.1@osu.edu}
         \affiliation{Department of Physics, The Ohio State
           University, Columbus, OH 43210, USA}

\author{Yossathorn Tawabutr}
         \email[Email: ]{tawabutr.1@osu.edu}
         \affiliation{Department of Physics, The Ohio State
           University, Columbus, OH 43210, USA}

\begin{abstract}
We construct a numerical solution of the recently-derived large-$N_c \& N_f$ small-$x$ helicity evolution equations \cite{Kovchegov:2015pbl} with the aim to establish the small-$x$ asymptotics of the quark helicity distribution beyond the large-$N_c$ limit explored previously in the same framework.     (Here $N_c$ and $N_f$ are the numbers of quark colors and flavors.) While the large-$N_c$ helicity evolution involves gluons only, the large-$N_c \& N_f$ evolution includes contributions from quarks as well. We find that adding quarks to the evolution makes quark helicity distribution oscillate as a function of $x$. Our numerical results in the large-$N_c \& N_f$ limit lead to the $x$-dependence of the flavor-singlet quark helicity distribution which is well-approximated by
\begin{align}\label{DSigmaMain}
\Delta \Sigma (x, Q^2)\bigg|_{\mbox{large-}N_c \& N_f} \sim \left( \frac{1}{x} \right)^{\alpha_h^q} \, \cos \left[ \omega_q \, \ln \left( \frac{1}{x} \right) + \varphi_q \right].  
\end{align} 
The power $\alpha_h^q$ exhibits a weak $N_f$-dependence, and, for all $N_f$ values considered, remains very close to $\alpha_h^q (N_f=0) = (4/\sqrt{3}) \sqrt{\alpha_s N_c/(2 \pi)}$  obtained earlier in the large-$N_c$ limit \cite{Kovchegov:2016weo,Kovchegov:2017jxc}. The novel oscillation frequency $\omega_q$ and phase shift $\varphi_q$ depend more strongly on the number of flavors $N_f$ (with $\omega_q =0$ in the pure-glue large-$N_c$ limit). The typical period of oscillations for $\Delta \Sigma$ is rather long, spanning many units of rapidity. We speculate whether the oscillations we find are related to the sign variation with $x$ seen in the strange quark helicity distribution extracted from the data \cite{deFlorian:2009vb,Ball:2013tyh,deFlorian:2014yva,Ethier:2017zbq}.  
\end{abstract}

\pacs{12.38.-t, 12.38.Bx, 12.38.Cy}

\maketitle
\tableofcontents


\section{\label{sec:intro}Introduction}

Understanding the partonic (quark and gluon) structure of the proton is an essential part of our understanding of Quantum Chromodynamics (QCD). One of the big open questions in the studies of proton structure is the proton spin puzzle. Proton is a composite spin-$1/2$ particle made of quarks and gluons. Consequently, the spin of the proton should be a sum of all the spins and orbital angular momenta (OAM) carried by the quarks and gluons comprising the proton. This statement is formalized in terms of helicity sum rules, one due to Jaffe and Manohar \cite{Jaffe:1989jz} and another one due to Ji \cite{Ji:1996ek}. The former reads
\begin{equation}
S_q+L_q+S_G+L_G=\frac{1}{2}
\label{eqn:JM}
\end{equation}
where $S_q$ and $S_G$ are the total spin carried by the quarks and gluons, respectively, whereas $L_q$ and $L_G$ are the OAM carried by the quarks and gluons in the proton.

One can write $S_q$ and $S_G$, which are functions of the momentum scale $Q^2$, as integrals of helicity distribution functions over the Bjorken $x$ variable,
\begin{align}
S_q(Q^2) = \frac{1}{2} \int\limits_0^1 dx \; \Delta\Sigma(x,Q^2), \ \ \  \ \
S_G(Q^2) = \int\limits_0^1 dx \; \Delta G(x,Q^2),
\label{eqn:SqSG}
\end{align}
where
\begin{equation}
\Delta\Sigma(x,Q^2) = \sum_{f=u,d,s,\ldots} \left[\Delta f(x,Q^2) + \Delta\bar{f}(x,Q^2)\right]
\label{eqn:DeltaSigma}
\end{equation}
is the flavor-singlet quark helicity distribution function. Here $\Delta f$ and $\Delta\bar{f}$ denote the helicity distribution functions of quarks and antiquarks, respectively. In \eq{eqn:SqSG}, $\Delta G$ is the gluon helicity distribution. The quark and gluon OAM, $L_q$ and $L_G$, can also be written as $x$-integrals of the OAM distributions \cite{Bashinsky:1998if,Hagler:1998kg,Harindranath:1998ve,Hatta:2012cs,Ji:2012ba}. Note that the latter cannot be expressed as expectation values of twist-2 operators in the proton state, unlike the helicity parton distribution functions (hPDFs).  

The proton spin puzzle started with the discrepancy between the prediction of the (perhaps, na\"{i}ve) constituent quark model stating that all the proton spin is carried by the constituent quarks (such that $S_q = 1/2$) and the experimental measurements by the European Muon Collaboration (EMC) \cite{Ashman:1987hv,Ashman:1989ig} that reported a much lower number for $S_q$. More recently, the experiments at the Relativistic Heavy Ion Collider (RHIC) measured a non-zero gluon spin $S_G$ \cite{Adamczyk:2014ozi,Adare:2015ozj}. The current experimental values for quark and gluon spin are $S_q (Q^2 = 10\, \mbox{GeV}^2) \approx 0.15 \div 0.20$, integrated over $0.001 < x <1$, and $S_G (Q^2 = 10 \, \mbox{GeV}^2) \approx 0.13 \div 0.26$, integrated over $0.05 < x <1$ (see \cite{Accardi:2012qut,Leader:2013jra,Aschenauer:2013woa,Aschenauer:2015eha,Aidala:2020mzt} for recent reviews). One can see that these reported $S_q$ and $S_G$ do not add up to 1/2 and the proton spin puzzle is still open. (Note also that none of the terms in \eq{eqn:JM} is positive-definite.) We conclude that the missing part of the proton's spin must come either from the smaller-$x$ regions in the integrals of \eqref{eqn:SqSG} for the spin terms $S_q$ and $S_G$ than the $x$-ranges reported above or from the OAM terms $L_q$ and $L_G$. 

No experiment, present or future, can measure helicity PDFs all the way down to $x=0$, since this would require infinite energy. In addition, at very small (but non-zero) values of $x$ higher-twist corrections become comparable to the leading-order contribution for many observables, making it hard (if not impossible) to extract twist-2 hPDFs from the data. (The way such corrections may come in is described using the parton saturation framework, see \cite{Iancu:2003xm,Weigert:2005us,JalilianMarian:2005jf,Gelis:2010nm,Albacete:2014fwa,Kovchegov:2012mbw} for reviews.) Therefore, to assess the amount of parton spin and OAM coming from the small-$x$ region, one has to develop a robust theoretical formalism, which, if able to describe the existing experimental data and predict the values of the future longitudinal spin measurements, can be extrapolated down to $x=0$ with a (hopefully) good level of confidence. 

First theoretical calculation of helicity distributions at small $x$ was carried out over two decades ago by Bartels, Ermolaev, and Ryskin \cite{Bartels:1995iu,Bartels:1996wc} using infrared evolution equations technique constructed by Kirschner and Lipatov~\cite{Kirschner:1983di}
(see also~\cite{Kirschner:1994rq,Kirschner:1994vc,Griffiths:1999dj}) to resum powers of the leading parameter $\as \ln^2(1/x)$ with $\as$ the strong coupling constant. Resummation of $\as \ln^2(1/x)$ is referred to as the double logarithmic approximation (DLA). The more recent years have seen renewed efforts to construct a formalism capable of describing helicity PDFs and OAM distributions at small-$x$ \cite{Kovchegov:2015pbl,Hatta:2016aoc,Chirilli:2018kkw,Kovchegov:2019rrz,Boussarie:2019icw}. In a series of papers \cite{Kovchegov:2015pbl,Kovchegov:2016zex,Kovchegov:2016weo,Kovchegov:2017jxc,Kovchegov:2017lsr,Kovchegov:2018znm,Cougoulic:2019aja} the helicity evolution at small-$x$ was constructed using the shock-wave formalism \cite{Balitsky:1995ub,Balitsky:1998ya,Kovchegov:1999yj,Kovchegov:1999ua,Weigert:2000gi} while employing the so-called polarized Wilson lines, similar to (but not the same as) the formalism employed in deriving the unpolarized Balitsky--Kovchegov (BK) \cite{Balitsky:1995ub,Balitsky:1998ya,Kovchegov:1999yj,Kovchegov:1999ua} and Jalilian-Marian--Iancu--McLerran--Weigert--Leonidov--Kovner
(JIMWLK)
\cite{Jalilian-Marian:1997dw,Jalilian-Marian:1997gr,Weigert:2000gi,Iancu:2001ad,Iancu:2000hn,Ferreiro:2001qy} evolution equations, which employed regular light-cone Wilson lines. 

Helicity evolution equations were derived in \cite{Kovchegov:2015pbl,Kovchegov:2016zex,Kovchegov:2018znm} using the DLA. Similar to the BK evolution, the equations close in the large-$N_c$ limit, with $N_c$ the number of quark colors. In addition, and different from the unpolarized case, the helicity evolution equations \cite{Kovchegov:2015pbl,Kovchegov:2016zex,Kovchegov:2018znm} also close in the large-$N_c \& N_f$ limit, where $N_f$ is the number of quark flavors. Owing to the complexity of the large-$N_c \& N_f$ equations \cite{Kovchegov:2015pbl,Kovchegov:2018znm}, only the large-$N_c$ equations have been solved to date, resulting in the following power-law small-$x$ asymptotics for the quark and gluon helicity distributions in DLA \cite{Kovchegov:2016weo,Kovchegov:2017jxc,Kovchegov:2017lsr}
\begin{align}\label{SigmaG_LargeNc}
\Delta \Sigma (x, Q^2) \bigg|_{\mbox{large-}N_c} \sim \left(\frac{1}{x}\right)^{\alpha_h^q},   \ \ \ \ \
       \Delta G (x, Q^2) \bigg|_{\mbox{large-}N_c} \sim \left(\frac{1}{x}\right)^{\alpha_h^G} 
\end{align}
with
\begin{align}
\alpha_h^q \bigg|_{\mbox{large-}N_c} = \frac{4}{\sqrt{3}} \, \sqrt{\frac{\as \, N_c}{2 \pi}} , \ \ \ \ \ \alpha_h^G \bigg|_{\mbox{large-}N_c} = \frac{13}{4 \sqrt{3}} \, \sqrt{\frac{\as
        \, N_c}{2 \pi}}.
\end{align}
(When writing equations like \eqref{SigmaG_LargeNc} we suppress the potentially non-trivial $Q^2$ dependence of the involved observables, along with the possible sub-dominant $x$-dependence in the prefactor, and concentrate on the leading $x$ dependence of the quantities.) The small-$x$ asymptotics of quark and gluon OAM distributions at large $N_c$ were found in \cite{Kovchegov:2019rrz} using the same formalism.

The goal of this work is to numerically solve the large-$N_c \& N_f$ helicity evolution equations derived in \cite{Kovchegov:2015pbl,Kovchegov:2018znm} and use the solution to assess possible corrections to the large-$N_c$ result \eqref{SigmaG_LargeNc} for the asymptotic behavior of $\Delta\Sigma (x,Q^2)$ at small $x$. The results can be (and are) combined with the values of $\Delta\Sigma$ extracted from the experiment \cite{deFlorian:2009vb,Ball:2013tyh,deFlorian:2014yva,Ethier:2017zbq} to extrapolate $\Delta\Sigma$ to smaller values of $x$ and estimate the total quark contribution to the proton spin $S_q$. In addition, our results can be employed to find the small-$x$ asymptotics of $\Delta G$ and the OAM distributions for quarks and gluons in the large-$N_c \& N_f$ limit, that is, beyond the large-$N_c$ expressions for $\Delta G$ in \eq{SigmaG_LargeNc} and for the OAM small-$x$ asymptotics derived in \cite{Kovchegov:2019rrz}.

The paper is structured as follows. In Section~\ref{sec:setup}, we rewrite and simplify the large-$N_c \& N_f$ helicity evolution equations from \cite{Kovchegov:2015pbl,Kovchegov:2018znm}. The equations are written in terms of the quark and gluon polarized dipole amplitudes, which are defined below as correlators of the polarized and regular light-cone Wilson lines. Our numerical algorithm is presented in Sec.~\ref{sec:recursion}, where we rewrite the equations in terms of the variables convenient for our numerical approach, discretize the equations and cast them in the form suitable for implementing our algorithm. The resulting numerical solutions of the large-$N_c \& N_f$ helicity evolution equations for $N_f = 2,3,6$ are presented in Sec.~\ref{sec:results}, see Figs.~\ref{fig:GA2D} and \ref{fig:GA0eta} there. The surprising new result, best visible in \fig{fig:GA0eta}, is that the quark and gluon polarized dipole amplitudes {\sl oscillate} as functions of energy (or of $\ln (1/x)$), changing the sign back and forth. These oscillations of dipole amplitudes, in turn, result in oscillations in $\Delta \Sigma (x, Q^2)$ as a function of $\ln (1/x)$, as demonstrated in \eq{eqn:DelSigmaAsymp} or, equivalently, \eq{DSigmaMain} above, and depicted in \fig{fig:sign_delsigma}. Such oscillating behavior was absent in the large-$N_c$ (and $N_f=0$) analyses carried out in the previous works \cite{Kovchegov:2016weo,Kovchegov:2017jxc,Kovchegov:2017lsr,Kovchegov:2019rrz}, which only saw a power-of-$1/x$ growth \eqref{SigmaG_LargeNc} of helicity distributions at small $x$. The oscillations of $\Delta \Sigma (x, Q^2)$ with $\ln (1/x)$ are the main qualitative result of this paper. 

The parameters $\alpha_h^q$, $\omega_q$, and $\varphi_q$ from \eq{DSigmaMain}, describing $\Delta \Sigma (x, Q^2)$ at small $x$ and at large-$N_c \& N_f$, are extracted from our numerical solution in Appendix~\ref{sec:analysis_DS} and are summarized in \eq{eqn:DelSigmaAsymp2}. In Sec.~\ref{sec:results} we also present a fit \eqref{omega_fit} for the $N_f$-dependence of the oscillation frequency $\omega_q$ which increases with $N_f$, meaning that that the oscillation period gets shorter and, therefore, the oscillations become more pronounced, in the regime where more quark flavors become dynamically relevant. In Sec.~\ref{sec:pheno} we follow the method from \cite{Kovchegov:2016weo} to construct a preliminary estimate of the impact of the asymptotic form for $\Delta \Sigma (x, Q^2)$ given by \eq{DSigmaMain} on the amount of spin carried by the quarks in the proton. The results are given in Figs.~\ref{fig:xdelsigma} and \ref{fig:delsigmaxmin}, similar to \cite{Kovchegov:2016weo} indicating a potential for a large correction. We conclude in Sec.~\ref{sec:conclusion} and speculate whether the oscillations we find in our solution are related to the oscillation of the strange quark hPDFs extracted from the experimental data in \cite{deFlorian:2009vb,Ball:2013tyh,deFlorian:2014yva,Ethier:2017zbq}.


\section{Helicity Evolution Equations at Large-$N_c \& N_f$}
\label{sec:setup}

At small $x$, in the DLA, the flavor-singlet quark helicity PDF can be written as \cite{Kovchegov:2015pbl,Kovchegov:2016zex,Kovchegov:2018znm} 
\begin{align}\label{DSigma}
\Delta \Sigma (x, Q^2) \equiv \sum_f \, [ \Delta q^f (x, Q^2 ) + \Delta {\bar q}^f (x, Q^2 ) ] = \frac{N_c \, N_f}{2 \pi^3} \, \int\limits_{\Lambda^2/s}^1 \frac{d z}{z}  \, \int\limits_\frac{1}{z \, s}^\frac{1}{z \, Q^2} \, \frac{d x_{10}^2}{x_{10}^2} \,  Q (x_{10}^2, z),
\end{align}
where $s \approx Q^2/x$ is the center-of-mass energy squared, $\Lambda$ is a transverse momentum scale characterizing the target before small-$x$ evolution, and the impact parameter-integrated polarized quark dipole amplitude is defined by
\begin{align}\label{Q_imp_int}
Q (x_{10}^2, z) = \int d^2 \left( \frac{{\un x}_1 + {\un x}_0}{2}\right) \, Q_{10} (z),
\end{align}
where the polarized quark dipole amplitude is
\begin{align}\label{Qdef}
Q_{10} (z) \equiv \frac{1}{2 \, N_c} \, \mbox{Re} \, \left\langle \mathcal{T} \, \tr \left[ V_{\un 0} \, \left( V_{\un 1}^{pol}\right)^\dagger \right] +\mathcal{T} \, \tr \left[ V_{\un 1}^{pol} \, V_{\un 0}^\dagger \right]  \right\rangle. 
\end{align}
The dipole consists of a quark and an anti-quark located at transverse positions ${\un x}_1$ and ${\un x}_0$ depending on the term in \eq{Qdef}. In our notation the transverse vectors are denoted by ${\un x} = (x^1, x^2)$ with ${\un x}_{ij} = {\un x}_i - {\un x}_j$, such that ${\un x}_{10} = {\un x}_1 - {\un x}_0$. In addition, $x_{ij}$ denotes the magnitude of the vector ${\un x}_{ij}$. The light-cone variables are defined by $x^\pm = (x^0 \pm x^3)/\sqrt{2}$. Our longitudinally polarized proton moves in the $x^+$ direction. The quantity $z$ is the fraction of some original ($x^-$-direction-moving) probe's momentum carried by the softest (anti-)quark in the dipole. The angle brackets in \eq{Qdef} denote the averaging in the proton wave function in the small-$x$/saturation sense \cite{Iancu:2003xm,Weigert:2005us,JalilianMarian:2005jf,Gelis:2010nm,Albacete:2014fwa,Kovchegov:2012mbw}, albeit now taking into account that the proton is longitudinally polarized.

The polarized dipole amplitude \eqref{Qdef} is defined in terms of the light-cone fundamental Wilson lines 
\begin{align}\label{Vdef}
  V_{\un{x}} [b^-, a^-] = \mathcal{P} \exp \left[ i g
    \int\limits_{a^-}^{b^-} d x^- \, A^+ (x^+=0, x^-, {\un x})
  \right]
\end{align}
and the so-called polarized Wilson lines $V^{pol}$ \cite{Kovchegov:2015pbl,Kovchegov:2016zex,Kovchegov:2017lsr,Kovchegov:2018znm}, consisting of regular light-cone Wilson lines along with the sub-eikonal helicity-dependent local operator insertion(s), 
\begin{align} 
  \label{eq:Vpol_all} 
  & V^{pol}_{\un x} = i g \, p_1^+ \,
  \int\limits_{-\infty}^\infty d x^- \, V_{\ul x} [+\infty, x^-] \: 
  F^{12} (x^-, \un{x}) \: V_{\ul x} [x^- , -\infty] \\ & - g^2 \, p_1^+
  \int\limits_{-\infty}^\infty d x_1^- \, \int\limits_{x_1^-}^\infty d x_2^- \, V_{\ul x} [+\infty, x_2^-] \:  t^b \, {\psi}_\beta (x_2^-, {\un x})  \, U_{\ul x}^{ba} [ x_2^-,  x_1^-] \left[ \frac{1}{2} \, \gamma^+ \, \gamma^5 \right]_{\alpha\beta} \, {\bar \psi}_\alpha (x_1^-, {\un x}) \, t^a  \: V_{\ul x} [x_1^- , -\infty]. \notag 
\end{align}
(Note that the definition \eqref{eq:Vpol_all} is different by an extra factor of $s$ in overall normalization as compared to that used in earlier works \cite{Kovchegov:2017lsr,Kovchegov:2018znm}.) We also use an abbreviated notation where $Q_{10} = Q_{{\un x}_1 , {\un x}_0}$, $V_{\un 1}^{pol}= V_{{\un x}_1}^{pol}$, etc. As usual, $\mathcal{T}$ in \eq{Qdef} denotes time ordering of the operators. In equations \eqref{Vdef} and \eqref{eq:Vpol_all}, $A^+$ is the eikonal gluon field of the shock wave (often referred to as the target, or the target proton), $F^{12}$ is the helicity-dependent sub-eikonal field-strength tensor component of the target gluon field (both in the fundamental representation), while $\psi$ and $\bar \psi$ are the (sub-eikonal) quark and anti-quark fields of the shock wave. In addition, $p_1^+$ is the large light-cone momentum of the parton in the shock wave generating those sub-eikonal fields. In \eq{eq:Vpol_all} we have also employed the adjoint light-cone Wilson line
\begin{align}
  U_{\un{x}} [b^-, a^-] = \mathcal{P} \exp \left[ i g
    \int\limits_{a^-}^{b^-} d x^- \, {\cal A}^+ (x^+=0, x^-, {\un x})
  \right].
\end{align}
The matrices $t^a$ and $t^b$ are the fundamental generators of SU($N_c$) with the indices $a,b = 1 , \ldots , N_c^2-1$.

The large-$N_c \& N_f$ helicity evolution equations we are about to study mix the polarized quark dipole amplitude \eqref{Qdef} with the polarized gluon dipole amplitude \cite{Kovchegov:2017lsr,Kovchegov:2018znm} \footnote{We would like to stress that while the polarized quark dipole amplitude $Q_{10}$ is related to the flavor-singlet quark hPDF $\Delta\Sigma$, the polarized gluon helicity amplitude $G_{10}$ is not related to the gluon hPDF $\Delta G$. The distribution $\Delta G$ at small-$x$ is related to a different type of the polarized dipole amplitude, as detailed in \cite{Kovchegov:2017lsr}.}
\begin{align}\label{eq:Gadj_def3}
G_{10} (z) = \frac{1}{2 (N_c^2 -1)} \, \mbox{Re} \, \left\langle \mathcal{T} \, \mbox{Tr} \left[ U_{\ul 0} \, U_{{\un 1}}^{pol \, \dagger} \right] + \mathcal{T} \, \mbox{Tr} \left[ U_{{\un 1}}^{pol} \, U_{\ul 0}^\dagger \right] \right\rangle
\end{align}
defined in terms of the adjoint polarized Wilson lines \cite{Kovchegov:2018znm}
\begin{align} 
  \label{M:UpolFull}
  & (U_{\ul x}^{pol})^{ab} = 2 i \, g \, p_1^+
  \int\limits_{-\infty}^{+\infty} dx^- \: \left( U_{\ul{x}}[+\infty, x^-] \:
  {\cal F}^{12} (x^+ =0 , x^- , \ul{x}) \: U_{\ul{x}} [x^- , -\infty] \right)^{ab} \\ &  - g^2 \, p_1^+ \, \int\limits_{-\infty}^\infty d x_1^- \, \int\limits_{x_1^-}^\infty d x_2^- \, U^{aa'}_{\un x} [+\infty, x_2^-] \,  {\bar \psi} (x_2^-, {\un x}) \, t^{a'} \, V_{\un x} [x_2^-, x_1^-] \, \frac{1}{2} \, \gamma^+ \gamma_5 \, t^{b'} \,  \psi (x_1^-, {\un x}) \, U^{b'b}_{\un x} [x_1^-, -\infty] - c.c.  . \notag
\end{align}
(Note again an additional $s$ in the normalization of \eq{M:UpolFull} as compared to \cite{Kovchegov:2018znm}; in addition, $G_{10} (z)$ in \eq{eq:Gadj_def3} was labeled $G^{adj}$ in \cite{Kovchegov:2018znm}.) The impact parameter-integrated polarized gluon dipole amplitude is defined similar to \eq{Q_imp_int}
\begin{align}\label{G_imp_int}
G (x_{10}^2, z) = \int d^2 \left( \frac{{\un x}_1 + {\un x}_0}{2}\right) \, G_{10} (z).
\end{align}

\begin{figure}
\includegraphics[width= 0.8 \textwidth]{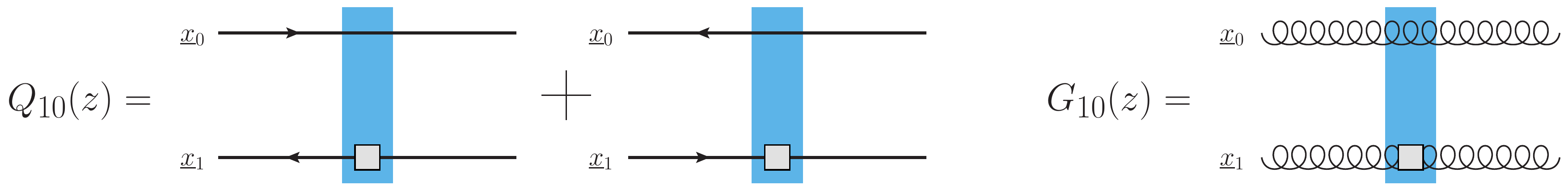} 
\caption{Diagrammatic representation of the quark ($Q$) and gluon ($G$) polarized dipole amplitudes. The shaded rectangle is the shock wave, while the square represents insertions of sub-eikonal operator(s).}
\label{fig:QG_diagrams}
\end{figure}

The polarized quark and gluon dipole amplitudes are depicted diagrammatically in \fig{fig:QG_diagrams}. The polarized proton, along with all the partons to be produced by the small-$x$ evolution, are depicted by the rectangular shock wave. The square on one of the lines depicts insertions of sub-eikonal operators present in Eqs.~\eqref{eq:Vpol_all} and \eqref{M:UpolFull}.

In the limit of large numbers of flavors, $N_f$, and colors, $N_c$, the small-$x$ DLA evolution equations for the polarized amplitudes $Q$ and $G$ are \cite{Kovchegov:2015pbl,Kovchegov:2018znm}
\begin{subequations}\label{eqn:ContEq1}
\begin{align}
G(x^2_{10},z) = G^{(0)}(x^2_{10},z) + & \ \frac{\alpha_sN_c}{2\pi}\int\limits_{\max\{\Lambda^2,1/x^2_{10}\}/s}^{z}\frac{dz'}{z'}\int\limits_{1/(z's)}^{x^2_{10}}\frac{dx^2_{21}}{x^2_{21}}\left[\Gamma(x^2_{10},x^2_{21},z')+3G(x^2_{21},z')\right] \label{Gevol} \\
&\;\;\;\;-\frac{\alpha_sN_f}{2\pi}\int\limits_{\Lambda^2/s}^{z}\frac{dz'}{z'}\int\limits_{1/(z's)}^{x^2_{10}z/z'}\frac{dx^2_{21}}{x^2_{21}}\;\overline{\Gamma}_{gen}(x^2_{10},x^2_{21},z'), \notag \\
Q(x^2_{10},z) =  Q^{(0)}(x^2_{10},z)+ & \frac{\alpha_sN_c}{4\pi}\int\limits_{\Lambda^2/s}^{z}\frac{dz'}{z'}\int\limits_{1/(z's)}^{x^2_{10}z/z'}\frac{dx^2_{21}}{x^2_{21}}\;Q(x^2_{21},z') \\
+ \frac{\alpha_sN_c}{2\pi}  \int\limits^{z}_{\max\{\Lambda^2,1/x^2_{10}\}/s} \frac{dz'}{z'} & \ \int\limits_{1/(z's)}^{x^2_{10}}\frac{dx^2_{21}}{x^2_{21}}\left[\frac{1}{2}G(x^2_{21},z')+\frac{1}{2}\Gamma(x^2_{10},x^2_{21},z')+Q(x^2_{21},z')-\overline{\Gamma}(x^2_{10},x^2_{21},z')\right]  ,  \notag  \\
\Gamma(x^2_{10},x^2_{21},z') = G^{(0)}(x^2_{10},z) +  & \ \frac{\alpha_sN_c}{2\pi}\int\limits_{\max\{\Lambda^2,1/x^2_{10}\}/s}^{z'}\frac{dz''}{z''}\int\limits_{1/(z''s)}^{\min\{x^2_{10},x^2_{21}z'/z''\}}\frac{dx^2_{32}}{x^2_{32}}\left[\Gamma(x^2_{10},x^2_{32},z'')+3G(x^2_{32},z'')\right] \label{Qevol}\\
&\;\;\;\;-\frac{\alpha_sN_f}{2\pi}\int\limits_{\Lambda^2/s}^{z'}\frac{dz''}{z''}\int\limits_{1/(z''s)}^{x^2_{21}z'/z''}\frac{dx^2_{32}}{x^2_{32}}\;\overline{\Gamma}_{gen}(x^2_{10},x^2_{32},z''), \notag \\
\overline{\Gamma}(x^2_{10},x^2_{21},z') =  Q^{(0)}(x^2_{10},z) + & \ \frac{\alpha_sN_c}{4\pi}\int\limits_{\Lambda^2/s}^{z'}\frac{dz''}{z''}\int\limits_{1/(z''s)}^{x^2_{21}z'/z''}\frac{dx^2_{32}}{x^2_{32}}\;Q(x^2_{32},z'') \\
+   \frac{\alpha_sN_c}{2\pi}\int\limits^{z'}_{\max\{\Lambda^2,1/x^2_{10}\}/s} \frac{dz''}{z''} & \int\limits_{1/(z''s)}^{\min\{x^2_{10},x^2_{21}z'/z''\}} \frac{dx^2_{32}}{x^2_{32}}\left[\frac{1}{2}G(x^2_{32},z'')+\frac{1}{2}\Gamma(x^2_{10},x^2_{32},z'')+Q(x^2_{32},z'')-\overline{\Gamma}(x^2_{10},x^2_{32},z'')\right], \notag
\end{align}
\end{subequations}
where the generalized quark dipole amplitude is defined as \cite{Kovchegov:2017lsr,Kovchegov:2018znm}
\begin{equation}
\overline{\Gamma}_{gen}(x^2_{10},x^2_{21},z') = \theta(x_{10}-x_{21})\overline{\Gamma}(x^2_{10},x^2_{21},z') + \theta(x_{21}-x_{10})Q(x^2_{21},z').
\label{eqn:ContEq2}
\end{equation}
To properly impose the light-cone time ordering, the evolution equations \eqref{eqn:ContEq1} employ two auxiliary functions, the gluon and quark polarized ``neighbor" dipole amplitudes, $\Gamma(x^2_{10},x^2_{21},z')$ for $G(x^2_{10},z)$ and $\overline{\Gamma}(x^2_{10},x^2_{21},z')$ for $Q(x^2_{10},z)$ \cite{Kovchegov:2015pbl,Kovchegov:2016zex,Kovchegov:2017lsr,Kovchegov:2018znm}. The operator definitions for $\overline{\Gamma}(x^2_{10},x^2_{21},z')$ and $\Gamma(x^2_{10},x^2_{21},z')$ are given by the same Eqs.~\eqref{Qdef} and \eqref{eq:Gadj_def3}, respectively (integrated over all impact parameters). However, implicit in those definitions is the $x^-$-light-cone life-time of the dipole, proportional to $z \, x_{10}^2$ \cite{Kovchegov:2015pbl,Iancu:2015vea,Ducloue:2019ezk,Cougoulic:2019aja}. For the ``neighbor" dipole amplitudes $\overline{\Gamma}(x^2_{10},x^2_{21},z')$ and $\Gamma(x^2_{10},x^2_{21},z')$ the life-time variable is proportional to $z' \, x_{21}^2$, different from $z' \, x_{10}^2$ one would expect based on their dipole size $x_{10}$. 

As follows from \eq{DSigma}, we only need to find the amplitude $Q(x^2_{10},z)$ in order to construct the quark helicity PDF. However, the evolution equations \eqref{eqn:ContEq1} mix all four involved amplitudes, and have to be solved for all the amplitudes in order to obtain $Q(x^2_{10},z)$. 

In this paper we are chiefly interested in the asymptotic behaviors of the dipole amplitudes at high energies. In \cite{Kovchegov:2016weo} it was shown that the high-energy asymptotics of helicity amplitudes at large-$N_c$ is largely independent of the initial conditions/inhomogeneous terms in the evolution, which only affect the overall prefactor in the asymptotic expression. (In addition, it is well-known that the small-$x$ asymptotics of the Balitsky--Fadin--Kuraev--Lipatov (BFKL) evolution \cite{Kuraev:1977fs,Balitsky:1978ic} depends on the initial conditions only in its prefactor as well.) Inspired by these examples of independence of the initial conditions, we assume it to be the case in the large-$N_c \& N_f$ approximation as well, and, for convenience, we take the inhomogeneous terms in Eqs.~\eqref{eqn:ContEq1} to be 
\begin{align}\label{init_cond}
G^{(0)}(x^2_{10},z) = Q^{(0)}(x^2_{10},z) = 1 ,
\end{align}
instead of using the proper Born-level initial condition \cite{Kovchegov:2016weo}.

In the regime of our interest, $\frac{1}{z's}\ll x^2_{21}\ll x^2_{10}$, the definition \eqref{eqn:ContEq2} implies that
\begin{subequations}\label{eqn:ContEq4}
\begin{align}
&\int\limits_{\Lambda^2/s}^{z'}\frac{dz''}{z''}\int\limits_{1/(z''s)}^{x^2_{21}z'/z''}\frac{dx^2_{32}}{x^2_{32}}\;\overline{\Gamma}_{gen}(x^2_{10},x^2_{32},z'')  \\
& =  \int\limits_{\max\{\Lambda^2,1/x^2_{10}\}/s}^{z'}\frac{dz''}{z''}\int\limits_{1/(z''s)}^{\min\{x^2_{10},x^2_{21}z'/z''\}}\frac{dx^2_{32}}{x^2_{32}}\;\overline{\Gamma}(x^2_{10},x^2_{32},z'') + \int\limits_{\Lambda^2/s}^{x^2_{21}z'/x^2_{10}}\frac{dz''}{z''}\int\limits_{\max\{x^2_{10},1/(z''s)\}}^{x^2_{21}z'/z''}\frac{dx^2_{32}}{x^2_{32}}\; Q(x^2_{32},z''), \notag \\
&\int\limits_{\Lambda^2/s}^{z}\frac{dz'}{z'}\int\limits_{1/(z's)}^{x^2_{10}z/z'}\frac{dx^2_{21}}{x^2_{21}}\;\overline{\Gamma}_{gen}(x^2_{10},x^2_{21},z') \\
& =  \int\limits_{\max\{\Lambda^2,1/x^2_{10}\}/s}^{z}\frac{dz'}{z'}\int\limits_{1/(z's)}^{x^2_{10}}\frac{dx^2_{21}}{x^2_{21}}\;\overline{\Gamma}(x^2_{10},x^2_{21},z') +\int\limits_{\Lambda^2/s}^{z}\frac{dz'}{z'}\int\limits_{\max\{x^2_{10},1/(z's)\}}^{x^2_{10}z/z'}\frac{dx^2_{21}}{x^2_{21}}\; Q(x^2_{21},z'), \notag
\end{align}
\end{subequations}
allowing us to rewrite the evolution equations \eqref{eqn:ContEq1} as 
\begin{subequations}\label{eqn:ContEq10}
\begin{align}
G(x^2_{10},z) = 1 + & \frac{\alpha_sN_c}{2\pi}\int\limits_{\max\{\Lambda^2,1/x^2_{10}\}/s}^{z}\frac{dz'}{z'}\int\limits_{1/(z's)}^{x^2_{10}}\frac{dx^2_{21}}{x^2_{21}}\left[\Gamma(x^2_{10},x^2_{21},z')+3G(x^2_{21},z') - \frac{N_f}{N_c} \, \overline{\Gamma}(x^2_{10},x^2_{21},z')  \right]  \\
& -\frac{\alpha_sN_f}{2\pi}\int\limits_{\Lambda^2/s}^{z}\frac{dz'}{z'}\int\limits_{\max\{x^2_{10},1/(z's)\}}^{x^2_{10}z/z'}\frac{dx^2_{21}}{x^2_{21}}\; Q(x^2_{21},z'), \notag \\
Q(x^2_{10},z) =  1+ & \frac{\alpha_sN_c}{4\pi}\int\limits_{\Lambda^2/s}^{z}\frac{dz'}{z'}\int\limits_{1/(z's)}^{x^2_{10}z/z'}\frac{dx^2_{21}}{x^2_{21}}\;Q(x^2_{21},z') \\
+  \frac{\alpha_sN_c}{2\pi} & \int\limits^{z}_{\max\{\Lambda^2,1/x^2_{10}\}/s}  \frac{dz'}{z'}  \int\limits_{1/(z's)}^{x^2_{10}}\frac{dx^2_{21}}{x^2_{21}}  \left[\frac{1}{2}G(x^2_{21},z')+\frac{1}{2}\Gamma(x^2_{10},x^2_{21},z')+Q(x^2_{21},z')-\overline{\Gamma}(x^2_{10},x^2_{21},z')\right], \notag \\
\Gamma(x^2_{10},x^2_{21},z') = 1 - & \frac{\alpha_sN_f}{2\pi}\int\limits_{\Lambda^2/s}^{x^2_{21}z'/x^2_{10}}\frac{dz''}{z''}\int\limits_{\max\{x^2_{10},1/(z''s)\}}^{x^2_{21}z'/z''}\frac{dx^2_{32}}{x^2_{32}}\; Q(x^2_{32},z'')\\
+  \frac{\alpha_sN_c}{2\pi}  & \!\!\! \int\limits_{\max\{\Lambda^2,1/x^2_{10}\}/s}^{z'} \!\! \frac{dz''}{z''}\int\limits_{1/(z''s)}^{\min\{x^2_{10},x^2_{21}z'/z''\}}\frac{dx^2_{32}}{x^2_{32}}\left[\Gamma(x^2_{10},x^2_{32},z'')+3G(x^2_{32},z'') - \frac{N_f}{N_c} \, \overline{\Gamma}(x^2_{10},x^2_{32},z'') \right], \notag \\
\overline{\Gamma}(x^2_{10},x^2_{21},z') =  1 + & \ \frac{\alpha_sN_c}{4\pi}\int\limits_{\Lambda^2/s}^{z'}\frac{dz''}{z''}\int\limits_{1/(z''s)}^{x^2_{21}z'/z''}\frac{dx^2_{32}}{x^2_{32}}\;Q(x^2_{32},z'')  \\
+   \frac{\alpha_sN_c}{2\pi} \int\limits^{z'}_{\max\{\Lambda^2,1/x^2_{10}\}/s} & \!\!\! \frac{dz''}{z''} \int\limits_{1/(z''s)}^{\min\{x^2_{10},x^2_{21}z'/z''\}} \frac{dx^2_{32}}{x^2_{32}}\left[\frac{1}{2}G(x^2_{32},z'')+\frac{1}{2}\Gamma(x^2_{10},x^2_{32},z'')+Q(x^2_{32},z'')-\overline{\Gamma}(x^2_{10},x^2_{32},z'')\right]. \notag 
\end{align}
\end{subequations}
Our aim is to solve this system of integral equations numerically. Its solution for $Q(x^2_{10},z)$, via \eq{DSigma}, would give us the quark helicity distribution at small Bjorken $x$ and in the regime where both the gluon and quark contributions to evolution are relevant (due to the large-$N_c \& N_f$ limit employed in deriving Eqs.~\eqref{eqn:ContEq10}).


\section{Numerical Solution: Discretization and Algorithm}
\label{sec:recursion}

To numerically evaluate the integrals in Eqs.~\eqref{eqn:ContEq10}, we first make the following changes of variables
\begin{equation}
\eta^{(n)} = \sqrt{\frac{\alpha_sN_c}{2\pi}}\ln\frac{z^{(n)}s}{\Lambda^2},  \ \ \
s_{kl} =  \sqrt{\frac{\alpha_sN_c}{2\pi}}\ln\frac{1}{x^2_{kl}\Lambda^2}\;,
\label{eqn:ContEq6}
\end{equation}
where $z^{(n)}$ can be $z$, $z'$ or $z''$ which relates to $\eta$, $\eta'$ or $\eta''$ respectively. In addition, $kl = 10, 21$ or $32$. In terms of these new variables, Eqs.~\eqref{eqn:ContEq10} can be re-written as
\begin{subequations}\label{eqn:ContEq7}
\begin{align}
G(s_{10},\eta) &= 1 + \int\limits_{\max\{0,s_{10}\}}^{\eta}d\eta'\int\limits^{\eta'}_{s_{10}}ds_{21} \left[\Gamma(s_{10},s_{21},\eta')+3G(s_{21},\eta') -\frac{N_f}{N_c} \;\overline{\Gamma}(s_{10},s_{21},\eta')\right] \\
&- \frac{N_f}{N_c}\int\limits_{0}^{\eta}d\eta'\int\limits^{\min\{s_{10},\eta'\}}_{s_{10}+\eta'-\eta}ds_{21}\;Q(s_{21},\eta'), \notag \\
Q(s_{10},\eta) &= 1 + \int\limits^{\eta}_{{\max\{0,s_{10}\}}} d\eta'\int\limits^{\eta'}_{s_{10}}ds_{21}\left[\frac{1}{2}G(s_{21},\eta') +\frac{1}{2}\Gamma(s_{10},s_{21},\eta')+Q(s_{21},\eta')-\overline{\Gamma}(s_{10},s_{21},\eta')\right]\\
&\;\;\;\;+\frac{1}{2}\int\limits_0^{\eta}d\eta'\int\limits^{\eta'}_{s_{10}+\eta'-\eta}ds_{21}\;Q(s_{21},\eta'), \notag \\
\Gamma(s_{10},s_{21},\eta') &= 1 + \int\limits_{\max\{0,s_{10}\}}^{\eta'}d\eta'' \int\limits^{\eta''}_{\max\{s_{10},s_{21}+\eta''-\eta'\}}ds_{32}\left[\Gamma(s_{10},s_{32},\eta'')+3G(s_{32},\eta'') -\frac{N_f}{N_c}  \, \overline{\Gamma}(s_{10},s_{32},\eta'') \right]\\
&\;\;\;\;-\frac{N_f}{N_c}\int\limits_{0}^{\eta'+s_{10}-s_{21}}d\eta''\int\limits^{\min\{s_{10},\eta''\}}_{s_{21}+\eta''-\eta'}ds_{32}\;Q(s_{32},\eta'') , \notag\\
\overline{\Gamma}(s_{10},s_{21},\eta') &=  1 + \int\limits^{\eta'}_{{\max\{0,s_{10}\}}} d\eta''\int\limits^{\eta''}_{\max\{s_{10},s_{21}+\eta''-\eta'\}} \!\!\!\!\! ds_{32} \left[\frac{1}{2}G(s_{32},\eta'')+\frac{1}{2}\Gamma(s_{10},s_{32},\eta'')+Q(s_{32},\eta'')-\overline{\Gamma}(s_{10},s_{32},\eta'')\right] \notag \\
&\;\;\;\;+\frac{1}{2}\int\limits_{0}^{\eta'}d\eta'' \int\limits^{\eta''}_{s_{21}+\eta''-\eta'}ds_{32}\;Q(s_{32},\eta''). 
\end{align}
\end{subequations}
Note that the integrals in equations \eqref{eqn:ContEq7} include the regions with $s_{21} <0, s_{32} <0$. This means that $\Lambda$ is not an infrared (IR) cutoff, but a perturbative transverse momentum scale characterizing the proton at the start of our evolution. 

Next, we discretize the resulting equations \eqref{eqn:ContEq7}. In particular, let $i^{(n)}$ and $j^{(n)}$ be the discretized version of $s_{kl}$ and $\eta^{(n)}$, with step sizes $\Delta s$ and $\Delta\eta$, respectively, such that $\eta^{(n)} = j^{(n)} \, \Delta \eta \equiv \eta_{j^{(n)}}$ along with $s_{10} = i  \, \Delta s \equiv s_i$, $s_{21} = i'  \, \Delta s \equiv s_{i'}$, and $s_{32} = i''  \, \Delta s \equiv s_{i''}$. The discretized amplitudes are defined as $G_{ij} \equiv G(s_i,\eta_j)$, $Q_{ij} \equiv Q(s_i,\eta_j)$,  $\Gamma_{ikj} \equiv \Gamma(s_i, s_k, \eta_j)$, and ${\overline \Gamma}_{ikj} \equiv {\overline \Gamma} (s_i, s_k, \eta_j)$. The equations \eqref{eqn:ContEq7} are self-contained over the following region in $(s_{10}, \eta)$-plane: $\eta \in [0, \eta_{max}]$, $\eta - \eta_{max} \le s_{10} \le \eta$, where $\eta_{max}$ is some arbitrary positive upper value of the $\eta$-range. This is the region where we will solve them numerically. Discretized Eqs.~\eqref{eqn:ContEq7} are
\begin{subequations}\label{eqn:DiscEq1}
\begin{align}
& G_{ij} = 1 + \Delta\eta\Delta s\left[\sum_{j'=\max\{0,i\}}^{j-1}\sum^{j'}_{i'=i}\left[\Gamma_{ii'j'}+3G_{i'j'}-\frac{N_f}{N_c}\overline{\Gamma}_{ii'j'}\right] - \frac{N_f}{N_c}\sum_{j'=0}^{j-1}\sum^{\min\{i,j'\}}_{i'=i+j'-j}Q_{i'j'}\right], \label{Gdisc1} \\
& Q_{ij} = 1 + \Delta\eta\Delta s\left[\sum^{j-1}_{j'=\max\{0,i\}} \sum^{j'}_{i'=i}\left[\frac{1}{2}G_{i'j'}+\frac{1}{2}\Gamma_{ii'j'}+Q_{i'j'}-\overline{\Gamma}_{ii'j'}\right] + \frac{1}{2}\sum_{j'=0}^{j-1}\sum^{j'}_{i'=i+j'-j}Q_{i'j'}\right], \\
&\Gamma_{ikj} = 1+\Delta\eta\Delta s\left[\sum_{j'=\max\{0,i\}}^{j-1}\sum^{j'}_{i'=\max\{i,k+j'-j\}}\left[\Gamma_{ii'j'}+3G_{i'j'}-\frac{N_f}{N_c}\overline{\Gamma}_{ii'j'}\right] - \frac{N_f}{N_c}\sum_{j'=0}^{i+j-k-1}\sum^{\min\{i,j'\}}_{i'=k+j'-j}Q_{i'j'}\right], \\
& \overline{\Gamma}_{ikj} =  1+\Delta\eta\Delta s\left[\sum^{j-1}_{j'=\max\{0,i\}}  \sum^{j'}_{i'=\max\{i,k+j'-j\}}\left[\frac{1}{2}G_{i'j'}+\frac{1}{2}\Gamma_{ii'j'}+Q_{i'j'}-\overline{\Gamma}_{ii'j'}\right] + \frac{1}{2}\sum_{j'=0}^{j-1}\sum^{j'}_{i'=k+j'-j}Q_{i'j'}\right] .
\end{align}
\end{subequations}

In principle, Eqs.~\eqref{eqn:DiscEq1} can already be solved numerically, similar to \cite{Kovchegov:2016weo,Kovchegov:2017lsr}. Instead we will simplify these equations which will allow us to implement a faster algorithm in the numerical solution. We replace $j$ by $j-1$ in each of the four equations in \eqref{eqn:DiscEq1}. For instance, in equation \eqref{Gdisc1} involving $G_{ij}$, this replacement gives
\begin{equation}
G_{i(j-1)} =  1 + \Delta\eta\Delta s\left[\sum_{j'=\max\{0,i\}}^{j-2}\sum^{j'}_{i'=i}\left[\Gamma_{ii'j'}+3G_{i'j'}-\frac{N_f}{N_c}\overline{\Gamma}_{ii'j'}\right]-\frac{N_f}{N_c}\sum_{j'=0}^{j-2}\sum^{\min\{i,j'\}}_{i'=i+j'-j+1}Q_{i'j'} \right] .
\label{eqn:DiscEq2}
\end{equation}
Comparing \eq{Gdisc1} to \eq{eqn:DiscEq2}, we obtain
\begin{equation}
G_{ij} = G_{i(j-1)} + \Delta\eta\Delta s\left[\sum^{j-1}_{i'=i}\left[\Gamma_{ii'(j-1)}+3G_{i'(j-1)}-\frac{N_f}{N_c}\overline{\Gamma}_{ii'(j-1)}\right] -\frac{N_f}{N_c}\sum_{j'=0}^{j-1}Q_{(i+j'-j)j'}  - \frac{N_f}{N_c}Q_{i(j-1)} \right].
\label{eqn:DiscEq3}
\end{equation}
Since $z s \ge 1/x_{10}^2$, we have $\eta \ge s_{10}$, which leads to $j \ge i$. For $i=j$ we have $Q_{j (j-1)} =1$ in the last term of \eq{eqn:DiscEq3}: this is determined by the initial conditions. Including this additional contribution from a single point in the $i,j$ grid does not significantly affect the numerical solution.
 
Writing each equation in \eqref{eqn:DiscEq1} in this recursive form allows for a numerical evaluation with one fewer layer of loops, resulting in much shorter computation time for smaller step sizes, $\Delta\eta$ and $\Delta s$, and for a larger $\eta$-range, defined by $\eta \in [0, \eta_{max}]$. In order to write down the recursive equations similar to Eq \eqref{eqn:DiscEq3} for $Q_{ij}$, $\Gamma_{ikj}$ and $\overline{\Gamma}_{ikj}$, notice that
\begin{equation}
\sum_{j'=\max\{0,i\}}^{j-1}\sum^{j'}_{i'=\max\{i,k+j'-j\}} \!\!\! \Gamma_{ii'j'} - \sum_{j'=\max\{0,i\}}^{j-2}\sum^{j'}_{i'=\max\{i,k+j'-j+1\}} \!\!\! \Gamma_{ii'j'} 
= \sum_{j'=\max\{i+j-k,0\}}^{j-1} \!\!\!  \Gamma_{i(k+j'-j)j'} + \sum^{j-1}_{i'=k}\Gamma_{ii'(j-1)}  ,
\label{eqn:DiscEq5}
\end{equation}
where we have used the fact that $i<k<j$, a consequence of the regime in which $\Gamma$ (and $\overline \Gamma$) are defined and employed, $1/{z' s} \ll x^2_{21}\ll x^2_{10}$, or, equivalently, $\eta' \gg s_{21} \gg s_{10}$. Equations \eqref{eqn:DiscEq1} and \eqref{eqn:DiscEq5} imply that (again for $i<k<j$)
\begin{subequations}\label{eqn:DiscEq6}
\begin{align}
Q_{ij}&= Q_{i(j-1)} + \Delta\eta\Delta s\left[\sum^{j-1}_{i'=i}\left[\frac{1}{2}G_{i'(j-1)}+\frac{1}{2}\Gamma_{ii'(j-1)}+\frac{3}{2}Q_{i'(j-1)}-\overline{\Gamma}_{ii'(j-1)}\right] + \frac{1}{2}\sum_{j'=0}^{j-1}Q_{(i+j'-j)j'} \right], \\
\Gamma_{ikj} &= \Gamma_{ik(j-1)} +    \Delta\eta\Delta s\Bigg[\sum_{i'=k}^{j-1}\left[\Gamma_{ii'(j-1)}+3G_{i'(j-1)}-\frac{N_f}{N_c}\overline{\Gamma}_{ii'(j-1)}\right] -\frac{N_f}{N_c}\sum_{j'=0}^{i+j-k-1}Q_{(k+j'-j)j'}-\frac{N_f}{N_c}Q_{i(i+j-k-1)} \notag  \\
&\;\;\;\;\;\;\;\;\;\;+ \sum_{j'=\max\{i+j-k,0\}}^{j-1}\left[\Gamma_{i(k+j'-j)j'}+3G_{(k+j'-j)j'}-\frac{N_f}{N_c}\overline{\Gamma}_{i(k+j'-j)j'}\right]\Bigg], \\
\overline{\Gamma}_{ikj} &=  \overline{\Gamma}_{ik(j-1)}+\Delta\eta\Delta s\Bigg[\sum^{j-1}_{i'=k}\left[\frac{1}{2}G_{i'(j-1)}+\frac{1}{2}\Gamma_{ii'(j-1)}+\frac{3}{2}Q_{i'(j-1)}-\overline{\Gamma}_{ii'(j-1)}\right]+\frac{1}{2}\sum_{j'=0}^{j-1}Q_{(k+j'-j)j'} \notag \\
&\;\;\;\;\;\;\;\;\;\;+ \sum_{j'=\max\{0,i+j-k\}}^{j-1}\left[\frac{1}{2}G_{(k+j'-j)j'}+\frac{1}{2}\Gamma_{i(k+j'-j)j'}+Q_{(k+j'-j)j'}-\overline{\Gamma}_{i(k+j'-j)j'}\right] \Bigg]\;. 
\end{align}
\end{subequations}

We solve Eqs.~\eqref{eqn:DiscEq3} and \eqref{eqn:DiscEq6} numerically in steps along the $\eta$-axis. To obtain the values for $0\leq\eta\leq\eta_{\max}$, we start from $\eta = \Delta\eta$, which is equivalent to $j=1$, at which we use Eqs.~\eqref{eqn:DiscEq3} and \eqref{eqn:DiscEq6} to determine $G_{i1}$ and $Q_{i1}$ for $1-\frac{\eta_{\max}}{\Delta\eta}\leq i\leq 1$, together with $\Gamma_{ik1}$ and $\overline{\Gamma}_{ik1}$ for  $1-\frac{\eta_{\max}}{\Delta\eta}\leq i \le k\leq 1$, assuming that $G_{i0}= Q_{i0} = \Gamma_{ik0} = \overline{\Gamma}_{ik0} =1$ are determined by the inhomogeneous term in Eqs.~\eqref{eqn:DiscEq1}. Afterward, we repeat the same steps by applying Eqs.~\eqref{eqn:DiscEq3} and \eqref{eqn:DiscEq6} for $j=2$, then $j=3$, and so on, until $j_{max} = \frac{\eta_{\max}}{\Delta\eta}$. At each $j$, we compute the dipole amplitudes $G_{ij}$ and $Q_{ij}$ for $i$ in the range $j-\frac{\eta_{\max}}{\Delta\eta}\leq i \le j$ and also find the amplitudes $\Gamma_{ikj}$ and $\overline{\Gamma}_{ikj}$ for all pairs of $i$ and $k$ satisfying $j-\frac{\eta_{\max}}{\Delta\eta}\leq i \le k \le j$. This process determines the values of $Q(s_{10},\eta)$ in the $\eta \in [0, \eta_{max}]$, $\eta - \eta_{max} \le s_{10} \le \eta$ region, which we will use below to determine the high energy asymptotics of the polarized quark dipole amplitude. We always take $\Delta s = \Delta\eta$ for simplicity.


\section{Numerical Solution: Results}
\label{sec:results}

\begin{figure}[t!]
	\centering
	\begin{subfigure}{.35\textwidth}
		\includegraphics[width=\textwidth]{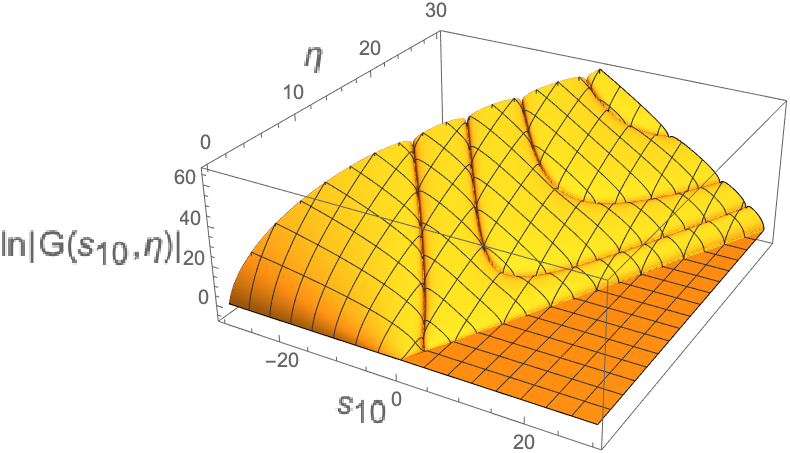}
		\caption{$\ln|G(s_{10},\eta)|$ at $N_f=2$}
	\end{subfigure}\;~\;~\;\;
	\begin{subfigure}{.35\textwidth}
		\includegraphics[width=\textwidth]{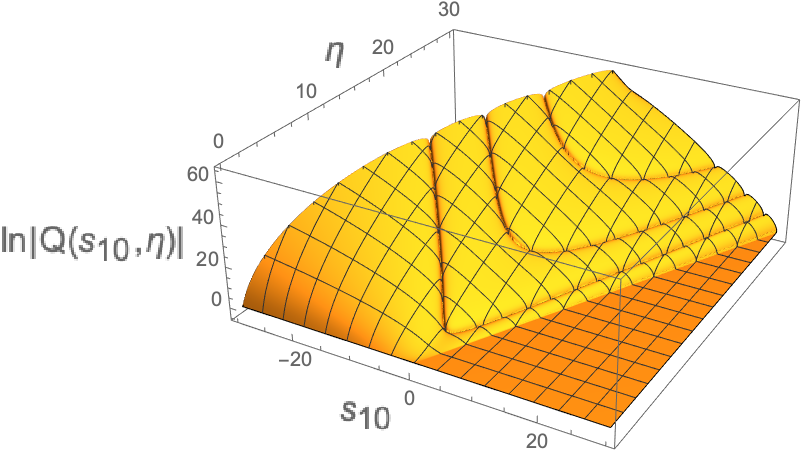}
		\caption{$\ln|Q(s_{10},\eta)|$ at $N_f=2$}
	\end{subfigure}\;~\;~\;\;

	\begin{subfigure}{.35\textwidth}
		\includegraphics[width=\textwidth]{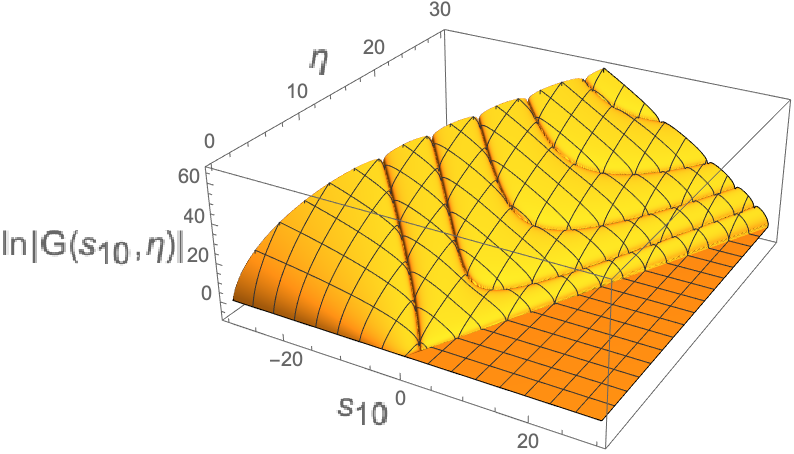}
		\caption{$\ln|G(s_{10},\eta)|$ at $N_f=3$}
	\end{subfigure}\;~\;~\;\;
	\begin{subfigure}{.35\textwidth}
		\includegraphics[width=\textwidth]{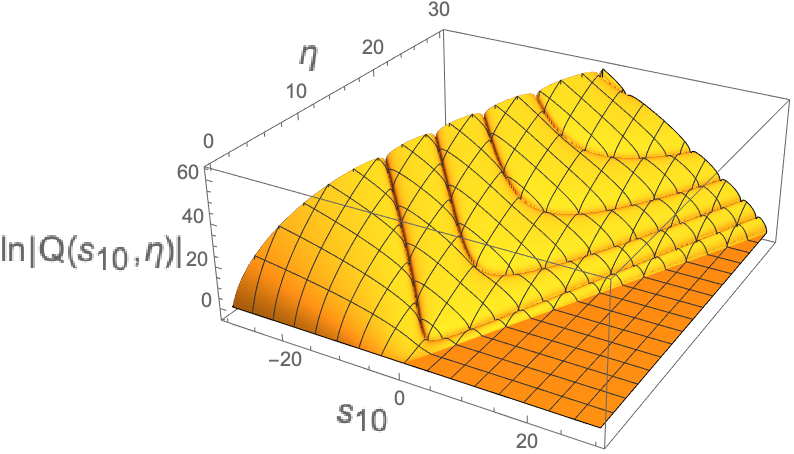}
		\caption{$\ln|Q(s_{10},\eta)|$ at $N_f=3$}
	\end{subfigure}\;~\;~\;\;

	\begin{subfigure}{.35\textwidth}
		\includegraphics[width=\textwidth]{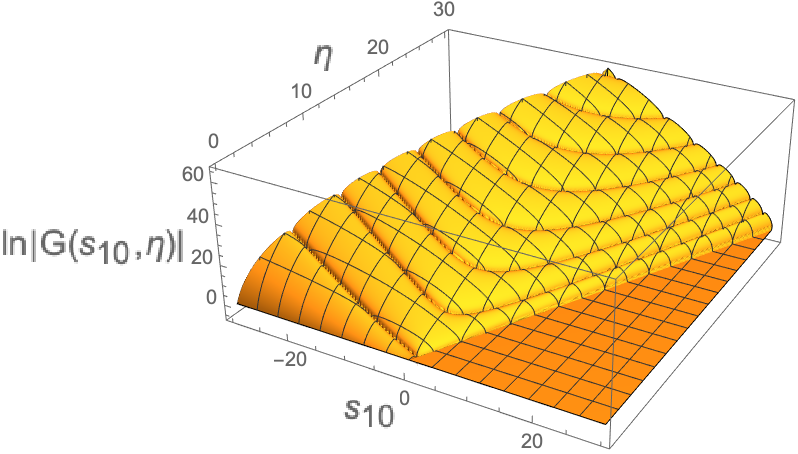}
		\caption{$\ln|G(s_{10},\eta)|$ at $N_f=6$}
	\end{subfigure}\;~\;~\;\;
	\begin{subfigure}{.35\textwidth}
		\includegraphics[width=\textwidth]{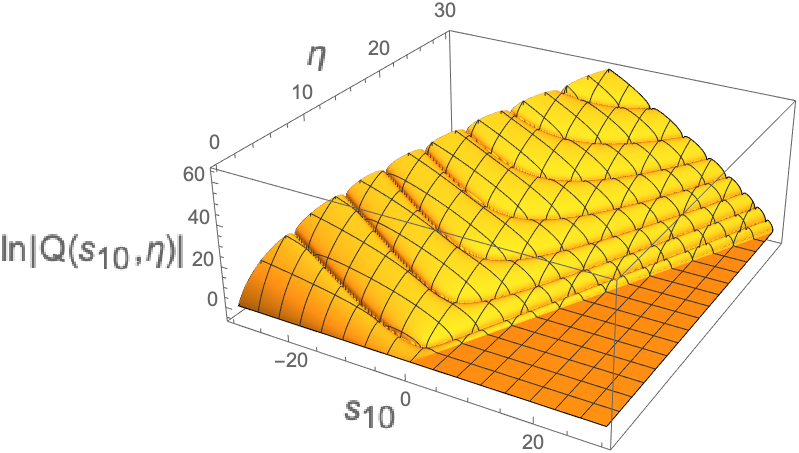}
		\caption{$\ln|Q(s_{10},\eta)|$ at $N_f=6$}
	\end{subfigure}
	\caption{Plots of $\ln|G(s_{10},\eta)|$ and $\ln|Q(s_{10},\eta)|$ for $N_f=2,3,6$ and $N_c =3$. All the graphs result from numerical computations with the step size $\Delta\eta = 0.075$ and $\eta_{\max} = 30$.}
\label{fig:GA2D}
\end{figure}

In this Section we present the results of our numerical solution of the large-$N_c \& N_f$ evolution equations \eqref{eqn:ContEq7} for the quark and gluon polarized dipole amplitudes $Q(s_{10},\eta)$ and $G(s_{10},\eta)$, along with the corresponding quark helicity PDF $\Delta\Sigma (x, Q^2)$. The plots of $\ln|G(s_{10},\eta)|$ and $\ln|Q(s_{10},\eta)|$ in the $(s_{10},\eta)$-plane are shown in \fig{fig:GA2D} for $N_f=2,3,6$. Throughout this paper, we take the number of colors to be $N_c=3$.

The plots in \fig{fig:GA2D} demonstrate an approximately linear rise of $\ln|G(s_{10},\eta)|$ and $\ln|Q(s_{10},\eta)|$ with $\eta$, similar to the large-$N_c$ case studied previously in \cite{Kovchegov:2016weo,Kovchegov:2017jxc}. The only difference is that now, in \fig{fig:GA2D}, the rise of those functions is not monotonic, and appears to be periodically interrupted by lines of sharp local minima.

\begin{figure}[h!]
	\centering
	\begin{subfigure}{.4\textwidth}
		\includegraphics[width=\textwidth]{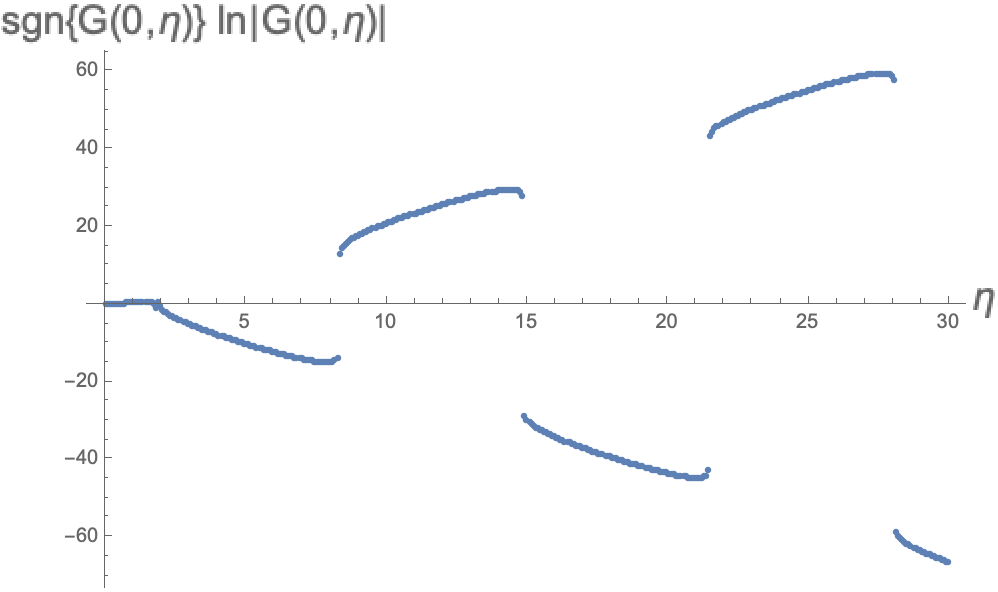}
		\caption{sgn$[G(0,\eta)] \, \ln|G(0,\eta)|$}
	\end{subfigure}\;\;\;\;\;\;
	\begin{subfigure}{.4\textwidth}
		\includegraphics[width=\textwidth]{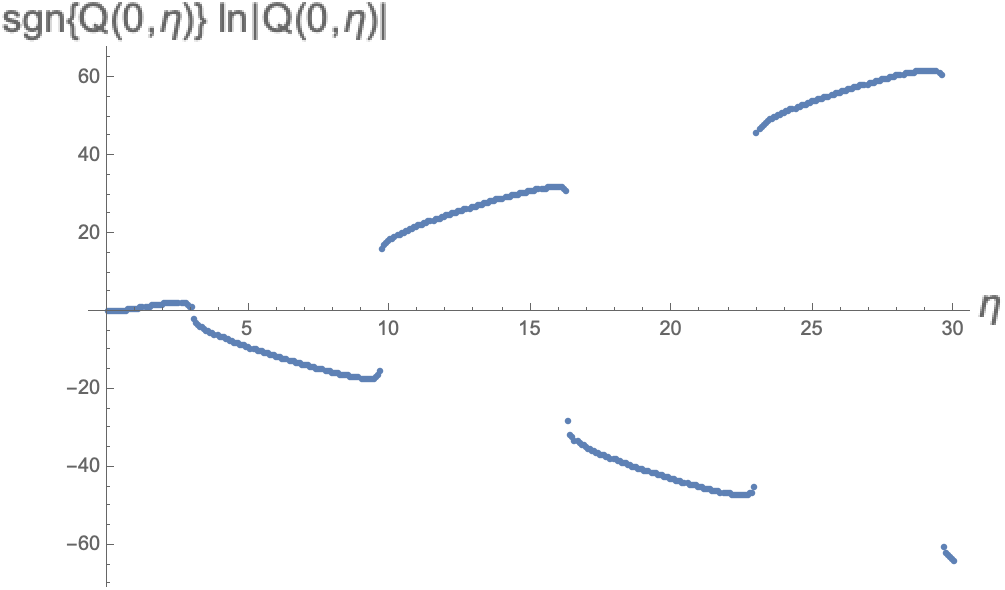}
		\caption{sgn$[Q(0,\eta)] \, \ln|Q(0,\eta)|$}
	\end{subfigure}
	\caption{Plots of sgn$[G(0,\eta)] \, \ln|G(0,\eta)|$ and sgn$[Q(0,\eta)] \,  \ln|Q(0,\eta)|$ versus $\eta$ for $N_f=3$ and $N_c =3$. Both graphs result from numerical computations with the step size $\Delta\eta = 0.075$ and $\eta_{\max} = 30$.}
\label{fig:GA0eta}
\end{figure}

To illustrate the origin of this non-monotonicity, we plot sgn$[G(0,\eta)] \, \ln|G(0,\eta)|$ and sgn$[Q(0,\eta)] \, \ln|Q(0,\eta)|$ as functions of $\eta$ in \fig{fig:GA0eta} for $N_f=3$. From these plots we see that $G(0,\eta)$ and $Q(0,\eta)$ {\sl oscillate with $\eta$}. The oscillations explain the non-monotonic behavior we saw in \fig{fig:GA2D}. These oscillations are the main qualitative difference of the small-$x$ asymptotics for the quark helicity distribution in the large-$N_c \& N_f$ limit, as compared to the large-$N_c$ case. It appears that introducing the quarks back into the helicity evolution equations generates oscillations. While the absolute values of $Q(s_{10},\eta)$ and $G(s_{10},\eta)$ still grow exponentially with $\eta$, both dipole amplitudes also oscillate. Moreover, from  \fig{fig:GA2D} one can see that the oscillation period appears to get smaller (and the oscillation frequency appears to get larger) with increasing $N_f$, which is also consistent with the fact that these oscillations are absent in the gluon-only large-$N_c$ equations solved in \cite{Kovchegov:2016weo,Kovchegov:2017jxc}. 

While an analytic solution of Eqs.~\eqref{eqn:ContEq7} is beyond the scope of this work, we can try to find analytic formulas approximating our numerical results in \fig{fig:GA0eta}, at least in the large-$\eta$ asymptotics. Combining the oscillations with the exponential growth of the maxima of $|Q(0,\eta)|$ and $|G(0,\eta)|$ with $\eta$ we propose the following asymptotic forms for the polarized dipole amplitudes:
\begin{subequations}\label{eqn:AsympNf1}
\begin{align}
& G(0,\eta) \sim e^{\alpha_G\eta}\cos\left(\omega_G\eta+\varphi_G\right) ,   \\
& Q(0,\eta) \sim e^{\alpha_Q\eta}\cos\left(\omega_Q\eta+\varphi_Q\right)    . \label{Qapp}
\end{align}
\end{subequations}
The oscillation frequencies are denoted by $\omega_G$ and $\omega_Q$, while the initial phases are denoted by $\varphi_G$ and $\varphi_Q$. As one can already see from \fig{fig:GA2D}, both the frequencies and the initial phases depend on $N_f$. This is confirmed by the detailed analysis of our numerical solution in Appendix~\ref{sec:analysis}. Furthermore, the amplitudes of oscillations in $Q(0,\eta)$ and $G(0,\eta)$ grow exponentially with $\eta$, while the exponents $\alpha_G$ and $\alpha_Q$ ({\sl the intercepts} in Regge terminology) also appear to depend on $N_f$. The results of the analysis carried out in Appendix~\ref{sec:analysis}, where we fit our numerical solution with the ansatz \eqref{eqn:AsympNf1} and extract the corresponding frequencies, phases, and intercepts, are summarized in Table~\ref{tab:aop}.

\begin{table}[h]
\begin{tabular}{|c|c|c|c|c|c|c|}
\hline
\multirow{2}{*}{$N_f$} & \multicolumn{3}{c|}{$G(0,\eta)$} & \multicolumn{3}{c|}{$Q(0,\eta)$} \\ \cline{2-7} 
                     & $\alpha_G$      & $\omega_G$    & $\varphi_G$      & $\alpha_Q$      & $\omega_Q$    & $\varphi_Q$     \\ \hline
2                   &     $2.321\pm 0.003$       &     $0.360 \pm 0.001$       &  $0.283\pm 0.011$ &    $2.320\pm 0.001$       &     $0.363\pm 0.002$   & $-0.452\pm 0.003$  \\ \hline
\;\;3\;\;           & \;\;$2.300\pm 0.004$\;\;   & \;\;$0.470\pm0.001$\;\;    &  $0.327 \pm 0.008$ &  \;\;$2.301\pm0.006$\;\;    &   \;\;$0.469\pm 0.005$\;\; & $-0.409\pm 0.025$   \\ \hline
6                   &    $2.291\pm 0.002$        &     $0.753\pm 0.004$       & \;\;$0.471\pm 0.009$\;\; &    $2.288\pm 0.002$      &     $0.753\pm 0.005$  & \;\;$-0.438\pm 0.012$\;\; \\ \hline
\end{tabular}
\caption{The resulting intercept $\alpha$, frequency $\omega$, and initial phase $\varphi$, for the gluon and quark polarized dipole amplitudes $G(0,\eta)$ and $Q(0,\eta)$ at $N_f=2,3,6$ and $N_c =3$. }
\label{tab:aop}
\end{table}

From Table~\ref{tab:aop} we see that, for each $N_f$, the two frequencies are equal within the numerical accuracy, $\omega_G = \omega_Q$. The frequencies increase with $N_f$: while the exact analytic dependence of $\omega_G = \omega_Q$ on $N_f$ is yet to be determined, we construct a Pad\'{e} approximant to write
\begin{align}\label{omega_fit}
\omega_Q = \omega_G  \approx \frac{0.22 N_f}{1 + 0.1265 N_f}  \approx \frac{0.07 \pi N_f}{1 + 0.1265 N_f}.
\end{align}

The intercepts, $\alpha_G$ and $\alpha_Q$, of the polarized dipole amplitudes' exponential growth given in Table~\ref{tab:aop}, are also equal to each other for each $N_f$ with the precision of our numerical solution, $\alpha_Q = \alpha_G$. For all $N_f$ studied they remain close to $\alpha_h^q (N_f=0) = \frac{4}{\sqrt{3}}\approx 2.309$, which is the intercept for $G(0,\eta)$ in the large-$N_c$ pure-glue limit with $N_f = 0$ (in units of $\sqrt{\as N_c/(2 \pi)}$), derived analytically in \cite{Kovchegov:2017jxc}. It appears, though, that $\alpha_Q = \alpha_G$ is decreasing slowly with $N_f$.

The initial phase $\varphi$ in Table~\ref{tab:aop} is always between $0$ and $\frac{\pi}{2}$ for $G(0,\eta)$ and between $-\frac{\pi}{2}$ and $0$ for $Q(0,\eta)$. The values of $\varphi_Q$ and $\varphi_G$ do not display a clear relation with each other. Their functional dependence on $N_f$ is non-monotonic, and its form is also not obvious from Table~\ref{tab:aop}. In fact, the initial phase depends greatly on the choice of the initial condition (the inhomogeneous term) in Eqs.~\eqref{eqn:ContEq1}. For instance, if one performs a similar computation using the Born-level-inspired dipole amplitudes as initial conditions (cf. \cite{Kovchegov:2016weo}), while still taking the inhomogeneous terms for $G$ and $Q$ to be equal for simplicity,
\begin{equation}
G^{(0)}(s_{10},\eta) = Q^{(0)}(s_{10},\eta) = \frac{\alpha^2_sC_F\pi}{2 \, N_c}\left(C_F\eta - 2(\eta-s_{10})\right),
\label{BornLvIC}
\end{equation} 
for $N_f = 3$, $N_c=3$, $\as = 0.35$, $\Delta\eta = 0.1$, $\eta_{\max} = 20$, and the fundamental Casimir operator $C_F = (N_c^2 -1)/2 N_c$ of SU($N_c$), one finds $\varphi_G = -1.146$ and $\varphi_Q = 1.530$, significantly different from all the initial phases displayed in Table~\ref{tab:aop}, which were obtained  for the initial condition \eqref{init_cond}. The difference between $\varphi_G = -1.146$ and $\varphi_Q = 1.530$ and the phases in Table~\ref{tab:aop} is too large to be attributed to discretization errors. We thus conclude that the phases $\varphi_Q$ and $\varphi_G$ are indeed very much dependent on the initial conditions to the evolution, and their values listed in Table~\ref{tab:aop} are not universal.

Next let us determine what our solution implies for the quark hPDF $\Delta\Sigma$. Rewriting \eq{DSigma} in terms of $\eta$ and $s_{10}$ from \eq{eqn:ContEq6} we arrive at
\begin{align}\label{DSigma2}
\Delta \Sigma (x, Q^2) = \frac{N_f}{\as \, \pi^2} \, \int\limits_{0}^{\sqrt{\frac{\as \, N_c}{2 \pi}} \ln \frac{Q^2}{x \Lambda^2}} d \eta  \, \int\limits_{\eta - \sqrt{\frac{\as \, N_c}{2 \pi}} \, \ln \frac{1}{x}}^\eta d s_{10} \  Q (s_{10}, \eta).
\end{align}
To determine the small-$x$ asymptotics of $\Delta \Sigma$ we need to evaluate the integral in \eq{DSigma2}. Note that for $Q \ge \Lambda$ the integration region in \eq{DSigma2} lies within the area $\eta \in [0, \eta_{max}]$, $\eta - \eta_{max} \le s_{10} \le \eta$ where we found the numerical solution for $Q (s_{10}, \eta)$ if we choose $\eta_{max} = \sqrt{\frac{\as \, N_c}{2 \pi}} \ln \frac{Q^2}{x \Lambda^2}$. Since $Q(s_{10},\eta)$ is known numerically, we perform a numerical integration to obtain the values of $\Delta\Sigma(x,Q^2)$ as a function of Bjorken $x$ at fixed $Q^2 = 10\text{ GeV}^2$ for the case of 3 quark flavors, $N_f=3$, while choosing, for simplicity, $\Lambda^2 = Q^2$, such that $\eta_{max} = \sqrt{\frac{\as \, N_c}{2 \pi}} \ln \frac{1}{x}$. 

\begin{figure}[h]
	\centering
		\includegraphics[width=0.6 \textwidth]{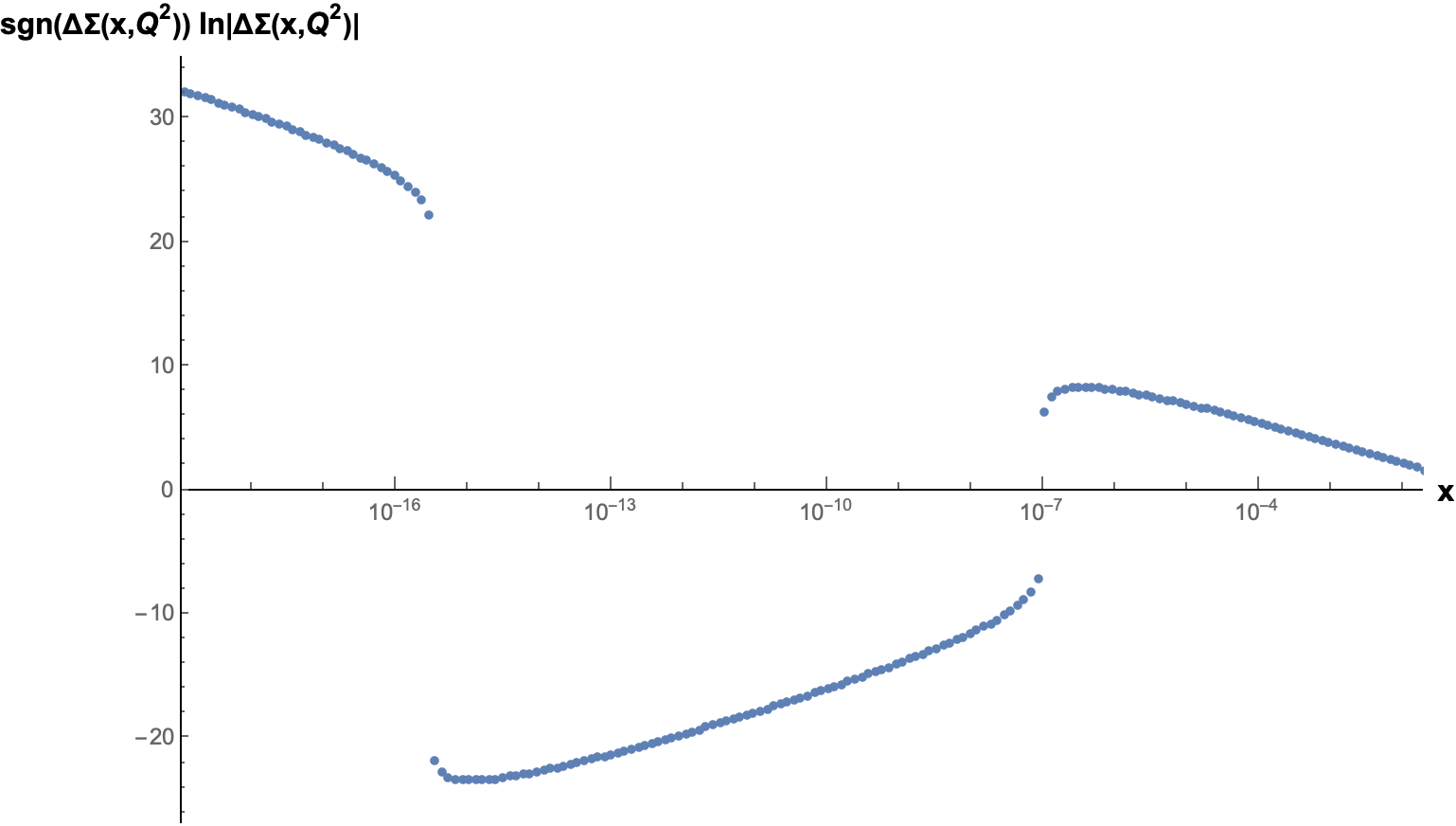}
	\caption{Plot of sgn$[\Delta\Sigma] \ln |\Delta\Sigma|$ versus $x$ resulting from the numerical integration of \eq{DSigma2} for $N_f=3$, $\as = 0.25$, $Q^2 = \Lambda^2$, and with the step size $\Delta\eta = 0.075$.}
\label{fig:sign_delsigma}
\end{figure}

The resulting $\Delta\Sigma(x,Q^2)$, similar to the polarized dipole amplitudes $Q$ and $G$, is an oscillating function of $\ln (1/x)$, with the oscillation amplitude growing exponentially with $\ln (1/x)$. In \fig{fig:sign_delsigma} we plot sgn$[\Delta\Sigma (x,Q^2)] \ln |\Delta\Sigma (x,Q^2)|$ as a function of $x$, demonstrating the oscillations explicitly. Inspired by the success of the ansatz \eqref{eqn:AsympNf1} for the dipole amplitude, we propose the following ansatz for the small-$x$ asymptotics of $\Delta\Sigma(x,Q^2)$,
\begin{align}\label{eqn:DelSigmaAsymp}
\Delta\Sigma(x,Q^2 = 10\text{ GeV}^2)\bigg|_{\mbox{large-}N_c \& N_f}  &\sim \left(\frac{1}{x}\right)^{\alpha_h^{q}}\cos\left[\omega_{q}\ln\frac{1}{x} + \varphi_{q}\right].
\end{align}
The parameters $\alpha_h^{q}$, $\omega_{q}$ and $\varphi_{q}$ are extracted from our numerical results in Appendix~\ref{sec:analysis_DS} by applying the parameter fitting process outlined in Appendix~\ref{sec:analysis}. We use $\as (10\text{ GeV}^2) \approx 0.25$. This gives (again, for $N_f=3$)
\begin{align}\label{eqn:DelSigmaAsymp2}
\alpha_h^{q} = (2.304\pm 0.012)\sqrt{\frac{\alpha_sN_c}{2\pi}}\;,\;\;\omega_{q} = (0.469 \pm 0.006)\sqrt{\frac{\alpha_sN_c}{2\pi}}\; , \; \;\text{and}\;\;\varphi_{q} = -1.25\pm 0.05.
\end{align}
The intercept and frequency are within the margins of error from those for the dipole amplitudes $Q$ and $G$ at $N_f=3$, multiplied by the factor of $\sqrt{\frac{\alpha_sN_c}{2\pi}}$: $\alpha_h^{q} = \alpha_Q \sqrt{\frac{\alpha_sN_c}{2\pi}} = \alpha_G \sqrt{\frac{\alpha_sN_c}{2\pi}}$, $\omega_{q} = \omega_{Q} \sqrt{\frac{\alpha_sN_c}{2\pi}} = \omega_G \sqrt{\frac{\alpha_sN_c}{2\pi}}$. This shows that the intercept and frequency of the dipole amplitude $Q$ determine the intercept and frequency of $\Delta\Sigma$: both are uniquely determined by the evolution. (Note that one cannot simply substitute \eq{Qapp} into \eq{DSigma2} to obtain this result analytically, since the former is valid only at $s_{10}=0$, while the latter has an integral over a range of non-zero values of $s_{10}$.) However, as remarked previously for the dipole amplitudes, the initial phase, $\varphi_{q}$, is, in fact, a by-product of our choice of initial conditions, $G^{(0)}$ and $Q^{(0)}$. In practical phenomenological applications, $G^{(0)}$ and $Q^{(0)}$, and, hence, the initial phase $\varphi_{q}$, have to be determined from the data.


\section{Quark Helicity: an Estimate}
\label{sec:pheno}

In this Section we follow the strategy employed in \cite{Kovchegov:2016weo} to estimate the possible impact of our new functional form for the small-$x$ asymptotics of $\Delta \Sigma(x,Q^2)$ \eqref{eqn:DelSigmaAsymp} on the amount of the proton spin carried by the small-$x$ quarks. Our analysis below should be understood as a rough estimate, with the anticipation of a more detailed phenomenology to be done in the future.

As the asymptotic form \eqref{eqn:DelSigmaAsymp} only holds for small $x$, we use that expression to extrapolate the quark helicity distribution from the DSSV14 result in \cite{deFlorian:2014yva}, starting at a particular connecting point, $x_0 \ll 1$, into the small-$x$ region, $x_{\min}\leq x\leq x_0$. In particular, we write the small-$x$ quark helicity distribution for $x< x_0$ as  
\begin{align}\label{eqn:DelSigmaAsymp3}
\Delta\Sigma(x,Q^2 = 10\text{ GeV}^2)\bigg|_{\mbox{large-}N_c \& N_f}  &= K\left(\frac{x_0}{x}\right)^{\alpha_h^{q}}\cos\left[\omega_{q}\ln\frac{x_0}{x} + \varphi_{q} \right],
\end{align}
with $\alpha_h^{q}$ and $\omega_{q}$ given in \eq{eqn:DelSigmaAsymp2}. For $x>x_0$ we take the quark hPDF to be given by the DSSV14 parameterization \cite{deFlorian:2014yva}, $\Delta\Sigma^{DSSV} (x,Q^2)$. We, therefore, need to match our $\Delta\Sigma(x,Q^2 = 10\text{ GeV}^2)$ from \eq{eqn:DelSigmaAsymp3} onto $\Delta\Sigma^{DSSV} (x,Q^2)$ at $x=x_0$. 

Unlike \cite{Kovchegov:2016weo}, where the ansatz for $\Delta\Sigma$ at small $x$ contained only the power of $1/x$, we now have the function in \eq{eqn:DelSigmaAsymp3} with two unknown parameters, the overall normalization factor $K$ and the phase $\varphi_{q}$. We assume that a more complete phenomenological approach would be able to uniquely determine $\varphi_{q}$ from the large-$x$ ($x>x_0$) data. We further assume that the value of $\varphi_{q}$ obtained from a more complete approach would generate a smooth matching of $\Delta\Sigma$ at $x=x_0$, ensuring continuity of both $\Delta\Sigma (x, Q^2)$ and its derivative $\pd \Delta\Sigma (x, Q^2)/ \pd x$ at $x=x_0$. This assumption is not very reliable for our purposes, since the functional form \eqref{eqn:DelSigmaAsymp3} is asymptotic, and is, therefore, valid only for $x \ll x_0$: using it to ensure the continuity of  $\Delta\Sigma (x, Q^2)$ and $\pd \Delta\Sigma (x, Q^2)/ \pd x$ at $x=x_0$ is somewhat questionable. Strictly-speaking we have to admit that $\varphi_{q}$ is an almost arbitrary parameter, whose value appears to be very important for assessing the amount of quark helicity at small $x$. While the more detailed phenomenology should better constraint the allowed ranges of $K$ and  $\varphi_{q}$, we will fix them here by simply requiring the continuity of  $\Delta\Sigma (x, Q^2)$ and $\pd \Delta\Sigma (x, Q^2)/ \pd x$ at $x=x_0$ between our asymptotics \eqref{eqn:DelSigmaAsymp3} and $\Delta\Sigma^{DSSV} (x,Q^2)$ from \cite{deFlorian:2014yva}. 

\begin{figure}[H]
	\centering
		\includegraphics[width=0.7 \textwidth]{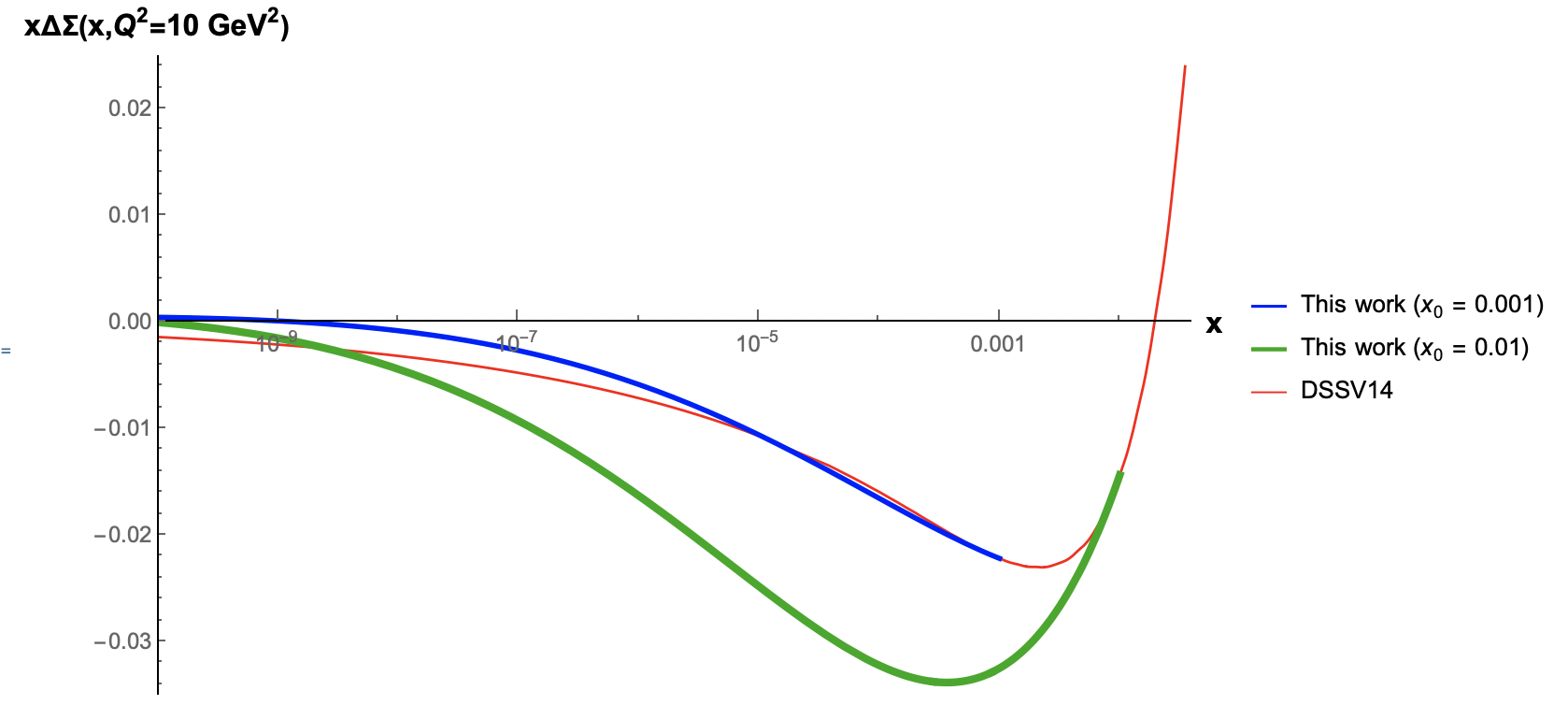}
	\caption{Plots of $x\Delta\Sigma(x,Q^2=10\text{ GeV}^2)$ versus $x$. The thin (red) line depicts the results from DSSV14 \cite{deFlorian:2014yva}, while the thick (green) and medium-thick (blue) lines show the results of matching our  asymptotics \eqref{eqn:DelSigmaAsymp3} to DSSV14 at $x_0 = 0.01$ and $x_0 = 0.001$, respectively.}
\label{fig:xdelsigma}
\end{figure}

We perform this computation for $x_0 = 0.01$ and $x_0 = 0.001$, doing a separate matching at each $x_0$. The resulting $x\Delta\Sigma(x,Q^2=10\text{ GeV}^2)$ is plotted versus $x$ in \fig{fig:xdelsigma}, which depicts the original DSSV14 \cite{deFlorian:2014yva} curve, extrapolated by a power-law in $x$ to very small $x$, along with the two curves resulting from matching our asymptotics \eqref{eqn:DelSigmaAsymp3} to DSSV14 at $x_0 = 0.01$ and $x_0 = 0.001$.

\begin{figure}[H]
	\centering
		\includegraphics[width=0.692 \textwidth]{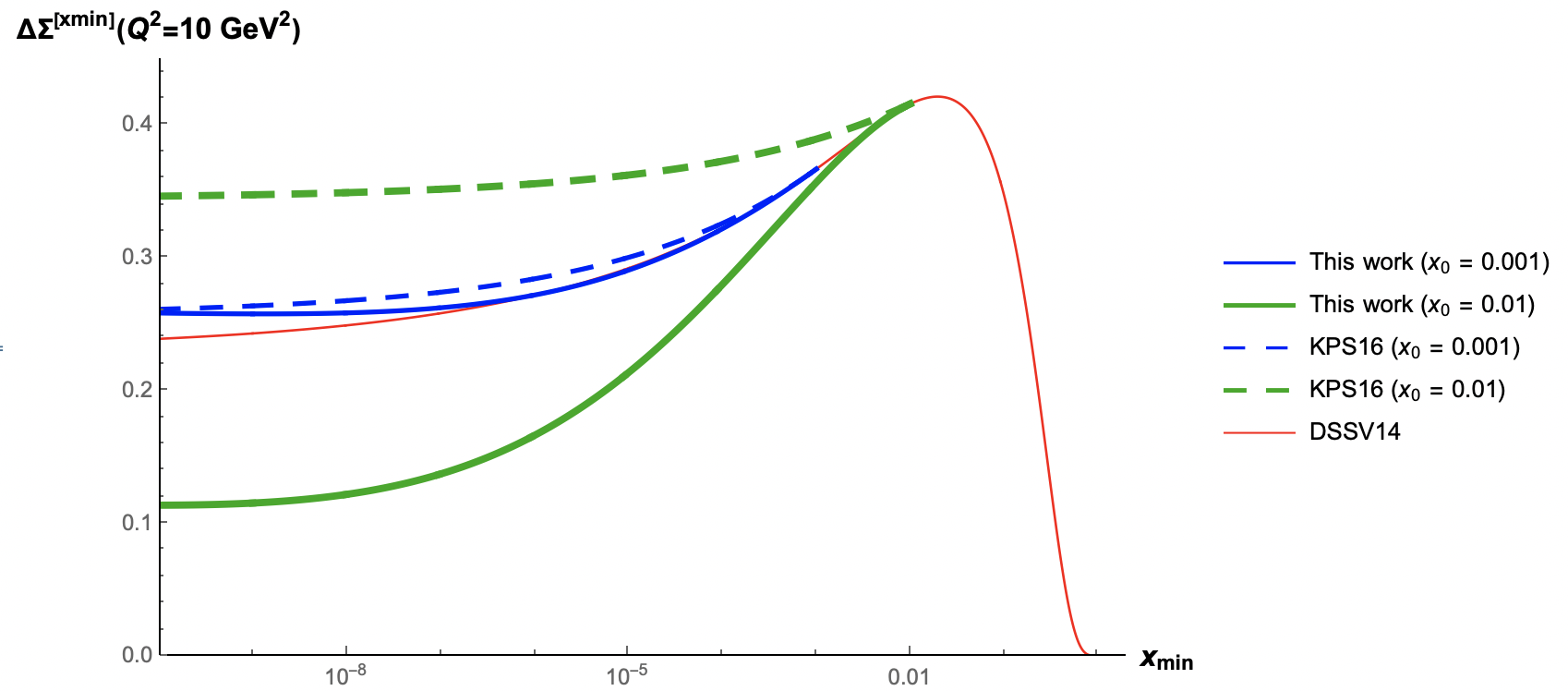}
	\caption{Plots of $\Delta\Sigma^{\left[x_{\min}\right]}(Q^2=10\text{ GeV}^2)$ defined in \eq{eqn:DelSigmaxmin} versus $x_{min}$. The thin (red) line is DSSV14 \cite{deFlorian:2014yva}, while the thick (green) and medium-thick (blue) solid lines show the results of matching our  asymptotics \eqref{eqn:DelSigmaAsymp3} to DSSV14 at $x_0 = 0.01$ and $x_0 = 0.001$, respectively. The two dashed lines represent the small-$x$ extrapolation done in \cite{Kovchegov:2016weo} using a power-law expression for $\Delta\Sigma (x, Q^2)$ at small $x$ resulting from the large-$N_c$ evolution, also matched onto DSSV14 at $x_0 = 0.01$ (thick dashed, green) and $x_0 = 0.001$ (medium-thick dashed, blue). }
\label{fig:delsigmaxmin}
\end{figure}

To see more clearly the implication of this result on the quark helicity, $S_q(Q^2 = 10\text{ GeV}^2)$, we consider the integral similar to \eq{eqn:SqSG} but with the lower limit at some small finite $x_{\min}$, such that $0<x_{\min}< x_0$ (cf. \cite{Kovchegov:2016weo})
\begin{align}\label{eqn:DelSigmaxmin}
\Delta\Sigma^{\left[x_{\min}\right]}(Q^2) = \int\limits_{x_{\min}}^{1}dx\;\Delta\Sigma(x,Q^2). 
\end{align}
From Eqs.~\eqref{eqn:SqSG} and \eqref{eqn:DelSigmaxmin} we see that $S_q(Q^2)= (1/2) \, \Delta\Sigma^{\left[x_{\min}\right]}(Q^2)$ in the $x_{\min}\to 0$ limit. In \fig{fig:delsigmaxmin}, we plot $\Delta\Sigma^{\left[x_{\min}\right]}(Q^2=10\text{ GeV}^2)$ versus $x_{\min}$ for the DSSV14 parameterization \cite{deFlorian:2014yva} along with the two curves resulting from matching our \eqref{eqn:DelSigmaAsymp3} to DSSV14 at $x_0 = 0.01$ and $x_0 = 0.001$. In addition, for comparison, we show two extra curves in \fig{fig:delsigmaxmin} (dashed lines, labeled KPS16), resulting from using the power-law ansatz for small-$x$ $\Delta\Sigma$ obtained from the large-$N_c$ pure-glue evolution \cite{Kovchegov:2016weo,Kovchegov:2017jxc}, and adjusting its overall normalization to ensure the continuity of $\Delta\Sigma (x, Q^2)$ with DSSV14 at $x_0 = 0.01$ and $x_0 = 0.001$, as was done in \cite{Kovchegov:2016weo}.

Figure~\ref{fig:delsigmaxmin} illustrates the potential of small-$x$ evolution to significantly affect the amount of the quark spin content in the proton, which was already observed in \cite{Kovchegov:2016weo}. We conclude from \fig{fig:delsigmaxmin} that the small-$x$ extrapolation of $\Delta\Sigma (x, Q^2)$ at $Q^2 = 10\text{ GeV}^2$ varies significantly with the matching point, $x_0$. For the larger $x_0$-value, $x_0 = 0.01$, we also see a strong variation between using large-$N_c$ (KPS16) and large-$N_c \& N_f$ small-$x$ evolution for helicity. It is also interesting to see in \fig{fig:delsigmaxmin} that for $x_0 = 0.001$ both the KPS16 curve and the curve based on \eq{eqn:DelSigmaAsymp3} are rather close to the DSSV14 curve and to each other. Again, let us stress that our use of the asymptotic expression \eqref{eqn:DelSigmaAsymp3}, valid for $x \ll x_0$, to perform the matching of both $\Delta\Sigma (x, Q^2)$ and $\pd \Delta\Sigma (x, Q^2)/ \pd x$ at $x=x_0$, which is probably outside of its region of applicability, gives us only a rough estimate of quark hPDF at small $x$. We expect that future more detailed studies would place this matching under a more solid theoretical control.


\section{Conclusions and Discussion}

\label{sec:conclusion}

In this work, we have numerically computed the asymptotic high-energy behavior of the polarized quark dipole amplitude resulting from the double-logarithmic small-$x$ helicity evolution \cite{Kovchegov:2015pbl,Kovchegov:2018znm} in the limit of large number of quark colors and flavors. The obtained amplitude $Q$ is plotted in Figs.~\ref{fig:GA2D} and \ref{fig:GA0eta} as a function of $\eta$ and $s_{10}$ defined in \eq{eqn:ContEq6}.  The amplitude $Q$ in the large-$N_c \& N_f$ limit displays an oscillatory pattern as a function of the center-of-mass energy and of the dipole transverse size, on top of the exponential growth seen before in the large-$N_c$ limit with $N_f=0$ \cite{Kovchegov:2016weo,Kovchegov:2017jxc}, when all the evolution was gluon-driven. We conclude that these oscillations appear after including quarks back into the small-$x$ helicity evolution of \cite{Kovchegov:2015pbl,Kovchegov:2018znm}. 

Our numerical results for $Q$ are well-approximated by \eq{Qapp}. The frequency $\omega_Q$ and initial phase $\varphi_Q$ of the oscillations depend on the number of flavors, with this dependence summarized in Table~\ref{tab:aop} and \eq{omega_fit}. The intercept $\alpha_Q$, also given in  Table~\ref{tab:aop}, exhibits weak dependence on $N_f$, largely staying close to the large-$N_c$, $N_f=0$ value $\alpha_h^q (N_f=0) = \frac{4}{\sqrt{3}}\approx 2.309$. The oscillatory feature of our solution can be attributed to the effect of the growing quark dipole amplitude counteracting the growing gluon amplitude, as manifested by the negative quark terms in Eqs.~\eqref{Gevol} and \eqref{Qevol}. To see in more detail how and why such counteraction leads to oscillations, it is desirable to construct an analytic solution of Eqs.~\eqref{eqn:ContEq1}. At this point, the complexity of Eqs.~\eqref{eqn:ContEq1} makes their analytic solution very hard to find. When and if such solution is constructed, the oscillations we observe would probably manifest themselves as complex-valued poles in Mellin space (conjugate to $\eta$). Let us also note that, while the periodic oscillations with energy we describe here are a new result for helicity distributions, a single sign reversal with decreasing $x$ was predicted for these quantities in \cite{Bartels:1996wc} using a different approach.

The polarized quark dipole amplitude oscillations result in the similar oscillating behavior of $\Delta \Sigma$ as a function of $\ln (1/x)$ with the oscillation amplitude growing as a power of $1/x$, as shown in \eq{DSigmaMain}, which we reproduce here
\begin{align}\label{DSigmaMain2}
\Delta \Sigma (x, Q^2)\bigg|_{\mbox{large-}N_c \& N_f} \sim \left( \frac{1}{x} \right)^{\alpha_h^q} \, \cos \left[ \omega_q \, \ln \left( \frac{1}{x} \right) + \varphi_q \right].  
\end{align}  
This is the main result of this work. It is illustrated in \fig{fig:sign_delsigma}. The intercept, $\alpha_h^q$, and the frequency, $\omega_q$, given by \eq{eqn:DelSigmaAsymp2}, are within the margin of error from the corresponding parameters, $\alpha_Q$ and $\omega_Q$, for the dipole amplitude, $Q$. The initial phase, $\varphi_q$, on the other hand, differs greatly from $\varphi_Q$, due to the relation \eqref{DSigma2} between the two quantities. In general, the phase $\varphi_Q$, and, therefore, the phase $\varphi_q$, exhibit a very strong dependence on the initial conditions (inhomogeneous terms) for our large-$N_c \& N_f$ evolution. This is in contrast to $\alpha_h^q \approx \alpha_Q \sqrt{\frac{\alpha_sN_c}{2\pi}}$ and $\omega_q \approx \omega_Q \sqrt{\frac{\alpha_sN_c}{2\pi}}$ which are independent of the initial conditions and are universal properties of the evolution. This ambiguity in the initial phase unfortunately trickles down to the ambiguity in our prediction for the quark helicity inside a proton. Fixing the phase should be done by determining the initial conditions for the evolution. Resolving this issue would require a more detailed phenomenological work, which is beyond the scope of this paper.

Note again that the oscillation frequency $\omega_q$ vanishes in the $N_f =0$ limit, such that the oscillation is a property of having quarks in the evolution. The  $N_f =0$, $N_c \to \infty$ limit itself warrants a little further discussion. In \cite{Kovchegov:2015pbl,Kovchegov:2016zex,Kovchegov:2016weo,Kovchegov:2017jxc,Kovchegov:2017lsr,Kovchegov:2018znm} the large-$N_c$ limit was understood as $N_f = 0$ gluons-only evolution. In this regime the polarized Wilson lines are given only by the first ($\sim F^{12}$) terms in Eqs.~\eqref{eq:Vpol_all} and \eqref{M:UpolFull}, such that $Q (x_{10}^2, z) \approx G (x_{10}^2, z) /4$ \cite{Kovchegov:2018znm}. The small-$x$ asymptotics of the quark and gluon amplitudes $Q$ and $G$ were, therefore, given by the same power of $1/x$. However, strictly-speaking one needs to clarify what is implied by the quark dipole amplitude at $N_f = 0$. The quark dipole amplitude obtained as a part of the gluon dipole amplitude in the large-$N_c$ limit, as employed in \cite{Kovchegov:2015pbl,Kovchegov:2016zex,Kovchegov:2016weo,Kovchegov:2017jxc,Kovchegov:2017lsr,Kovchegov:2018znm} and given by $Q (x_{10}^2, z) \approx G (x_{10}^2, z) /4$, is not exactly the right object to determine the quark helicity PDF $\Delta \Sigma$ in \eq{DSigma}. To properly define $\Delta \Sigma$ as the \emph{quark} helicity distribution in the large-$N_c$ limit, one needs to have a non-zero $N_f$. 
One may assume that the corresponding limit of small finite $N_f > 0$ and $N_c \to \infty$, with $\as \, N_c =$~const, can be imposed in equations \eqref{eqn:ContEq1} by dropping all the $N_f$-terms, that is, formally by putting $N_f=0$ in the equations. We report here that the solution of the resulting equations performed as part of this work leads to the intercept $\alpha_Q = 2.39$. If we compare this intercept to those listed in Table~\ref{tab:aop} we see that this result appears to support our earlier conclusion about mild $N_f$-dependence of the intercept. 

However, the large-$N_c$ finite-$N_f$ limit may need to be taken more carefully. Note that the small $N_f >0$, $N_c \to \infty$, $\as \, N_c =$~const limit is further complicated by the fact that in $Q (x_{10}^2, z)$ the Born-level interaction with the quark target, which should be used for calculating $Q^{(0)} (x_{10}^2, z)$, consists of a sum of two terms with different $N_c$-scaling \cite{Kovchegov:2016zex} (see \eq{BornLvIC} above): the $t$-channel quarks exchange is $N_c$-enhanced compared to the $t$-channel gluon exchange contribution. In general, the $N_c$ scaling of the terms in $Q^{(0)}$ and $G^{(0)}$ depends on whether the interaction happens with a quark or with a gluon in the target. We see that the small $N_f >0$, $N_c \to \infty$, $\as \, N_c =$~const limit should be taken by systematically performing the $1/N_c$ expansion in both the evolution equations \eqref{eqn:ContEq1} (or, more precisely, in the equations for Wilson lines from which Eqs.~\eqref{eqn:ContEq1} were obtained in \cite{Kovchegov:2015pbl,Kovchegov:2018znm}) and in the inhomogeneous terms $Q^{(0)}$ and $G^{(0)}$.  A careful implementation of this expansion shows that imposing the aforementioned limit by putting $N_f =0$ in equations \eqref{eqn:ContEq1} is correct only for a pure-glue shock wave, that is, for probing a polarized glueball state instead of the proton. For the scattering on the actual proton, or on a single polarized quark, the limit is more subtle, and requires a dedicated study. (It appears that, in this case, at leading-$N_c$ (and finite small $N_f$) one has to discard the gluon dipole amplitudes $G$ and $\Gamma$ from Eqs.~\eqref{eqn:ContEq1}, which is likely to result in a solution for $Q$ which is rather slowly-growing with energy. At the first subleading-$N_c$ order, one would get a sum of the solution of the Eqs.~\eqref{eqn:ContEq1} with $N_f =0$ but with $N_c$-suppressed part of the inhomogeneous terms, and the solution of Eqs.~\eqref{eqn:ContEq1} with leading-$N_c$ initial conditions and exactly one iteration of the $N_f$-terms.) This regime has not been explored in this work due to the higher phenomenological relevance of the large-$N_c \& N_f$ approximation considered here. 

 Finally, let us comment on the potential phenomenological implications of our main qualitative result: at small-$x$ the flavor-singlet quark helicity distribution $\Delta \Sigma (x, Q^2)$ oscillates in $\ln (1/x)$. Similar oscillation has been found in the strange quark helicity distribution $\Delta s$ extracted from the experimental data by the PDF collaborations \cite{deFlorian:2009vb,Ball:2013tyh,deFlorian:2014yva,Ethier:2017zbq}. This oscillation is the driving force behind the sign change of $\Delta \Sigma^{DSSV} (x, Q^2)$ in \fig{fig:xdelsigma} above. If the $\Delta s$ oscillation in $x$ is confirmed by the future data extractions, it appears reasonable to ask a question whether it is related to our oscillating result \eqref{DSigmaMain2} for $\Delta \Sigma (x, Q^2)$. While we do not separately consider individual quark flavors, the frequency of $\Delta \Sigma$ oscillations increases with $N_f$, and should be more pronounced if more flavors are included in the helicity evolution. This may be related to the oscillation observed in $\Delta s$, and not in $\Delta u$ or $\Delta d$. On the other hand the period of these oscillations in $\ln (1/x)$ is
\begin{align}
T = \frac{2 \pi}{\omega_q}\;.
\end{align}
Using $\omega_q =  (0.469)\sqrt{\frac{\alpha_sN_c}{2\pi}}$ from \eq{eqn:DelSigmaAsymp2} with $\alpha_s = 0.3$ and $N_c=3$, we obtain $T (N_f =3) \approx 35$, which is a very large number for rapidity or for $\ln (1/x)$. However, the first sign flip one encounters depends on the initial phase of the oscillation, which is hard to determine. If we start from the maximum of the cosine function in \eq{DSigmaMain2} at some $x_0$, then the first sign flip would happen at $x/x_0 \approx e^{-T/4} \approx 10^{-4}$, which is a more phenomenologically-reasonable number, but is still rather low to be relevant for the upcoming Electron-Ion Collider (EIC) experimental program \cite{Accardi:2012qut,Boer:2011fh,Aidala:2020mzt}. On yet another hand, the period of oscillations found above may be significantly affected by the higher-order corrections in $\as$, even by the running of the coupling. In addition, we are encouraged by the similarity of the shapes between our curves and the DSSV14 line in \fig{fig:xdelsigma}, indicating that some of the physics behind the DSSV14 line might be accurately described by our small-$x$ evolution. We, therefore, leave the final verdict on the issue of the phenomenological relevance of the $\Delta \Sigma$ oscillations found in this work for the future investigations.


\section{Acknowledgment}

\label{sec:acknowledgement}

The authors would like to thank Mr.~Daniel Adamiak for providing his code to help us construct Figs.~\ref{fig:xdelsigma} and \ref{fig:delsigmaxmin}, which, in turn, was based on the code provided by Prof.~Daniel Pitonyak, to whom we are also grateful. We also thank Prof.~Pitonyak for a discussion of numerical simulations for helicity at small $x$. YK would like to thank Mr.~Mohammed Karaki for his work on this project in its very early stages. 

This material is based upon work supported by the U.S. Department of
Energy, Office of Science, Office of Nuclear Physics under Award
Number DE-SC0004286.



\appendix
\section{Analysis of the solution for the polarized dipole amplitudes}

\label{sec:analysis}

This Appendix describes how we fit our numerical solution for the polarized dipole amplitudes $Q(0,\eta)$ and $G(0,\eta)$ using the ansatz  \eqref{eqn:AsympNf1} and extract the corresponding intercepts, frequencies, and phases. 

Given the numerical values of $G(0,\eta)$ and $Q(0,\eta)$ for a set of $\eta$ values with discrete spacing, we utilize the following method to deduce the intercepts ($\alpha_G$ and $\alpha_Q$), frequencies ($\omega_G$ and $\omega_Q$), and initial phases ($\varphi_G$ and $\varphi_Q$) of the oscillations. Consider a numerical simulation for a function of the form
\begin{equation}
f(\eta) = Ke^{\alpha \eta}\cos\left(\omega \, \eta+ \varphi \right)
\label{eqn:Analysis1}
\end{equation}
with some constants, $\alpha$, $\omega$, $\varphi$, and $K$. This is the asymptotic form assumed in Eqs.~\eqref{eqn:AsympNf1} for $G(0,\eta)$ and $Q(0,\eta)$.  The second derivative of the logarithm of $|f(\eta)|$ is
\begin{align}
\frac{d^2}{d\eta^2}\ln\left|f(\eta)\right| &= \frac{d}{d\eta}\left[\alpha - \omega \, \tan\left(\omega \, \eta+ \varphi \right)\right] = -\frac{\omega^2}{\cos^2\left(\omega \, \eta+ \varphi \right)}\;.
\label{eqn:Analysis2}
\end{align}
A local maximum of this second derivative occurs when $\cos\left(\omega \, \eta+ \varphi \right) = \pm 1$, and hence the frequency $\omega$ can be found from the value of the numerically-obtained second derivative at the maximum,
\begin{align}\label{max_omega}
\max \left[ \frac{d^2}{d\eta^2}\ln\left|f(\eta)\right| \right] = - \omega^2,
\end{align}
where we also adopt a convention in which $\omega >0$. We use the largest-$\eta$ maximum available in our simulation to extract $\omega$ using \eq{max_omega}. (Indeed the extracted value of $\omega$ can be cross-checked by comparing $\pi/\omega$ to the spacing between the positions of the local maxima along the $\eta$-axis in the numerical solution.) The phase $\varphi$ can then be determined from the second derivative maximum condition $\omega \, \eta^* + \varphi = \pi \, n$, where $\eta^*$ is the numerically-extracted position of the same largest-$\eta$ maximum of the second derivative \eqref{eqn:Analysis2} and $n$ is an integer. The value of $n$ is adjusted so that $\varphi \in (-\pi, \pi]$, where the choice between $\varphi \in (0, \pi]$ and $\varphi \in (-\pi, 0]$ is done by making sure that the corresponding $f(\eta^*)$ given by \eq{eqn:Analysis1} is positive or negative (assuming that $K>0$) in agreement with the numerical value of the function $f(\eta)$ at $\eta = \eta^*$. 

\begin{figure}[h]
	\centering
	\begin{subfigure}{.4\textwidth}
		\includegraphics[width=\textwidth]{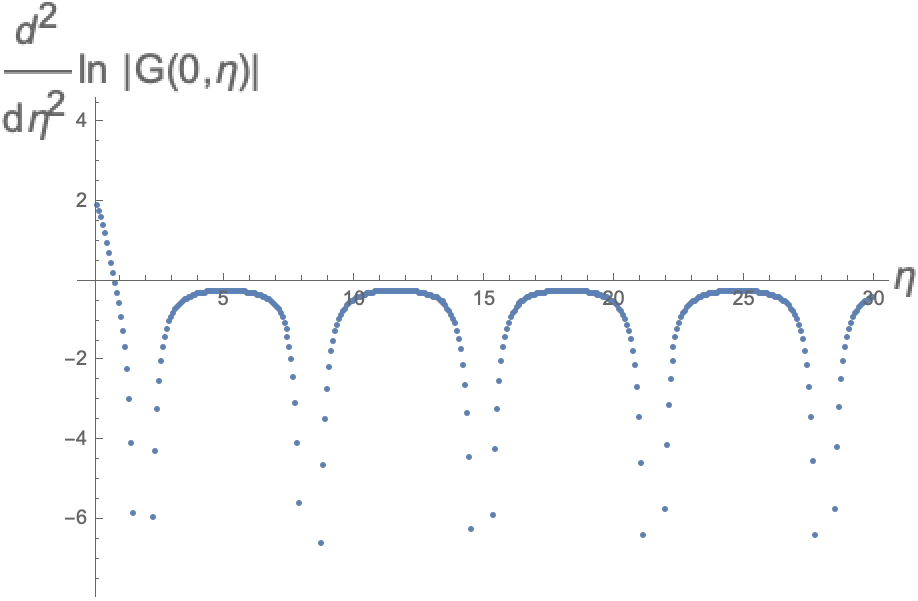}
		\caption{$\frac{d^2}{d\eta^2}\ln|G(0,\eta)|$}
	\end{subfigure}\;\;\;\;\;\;
	\begin{subfigure}{.4\textwidth}
		\includegraphics[width=\textwidth]{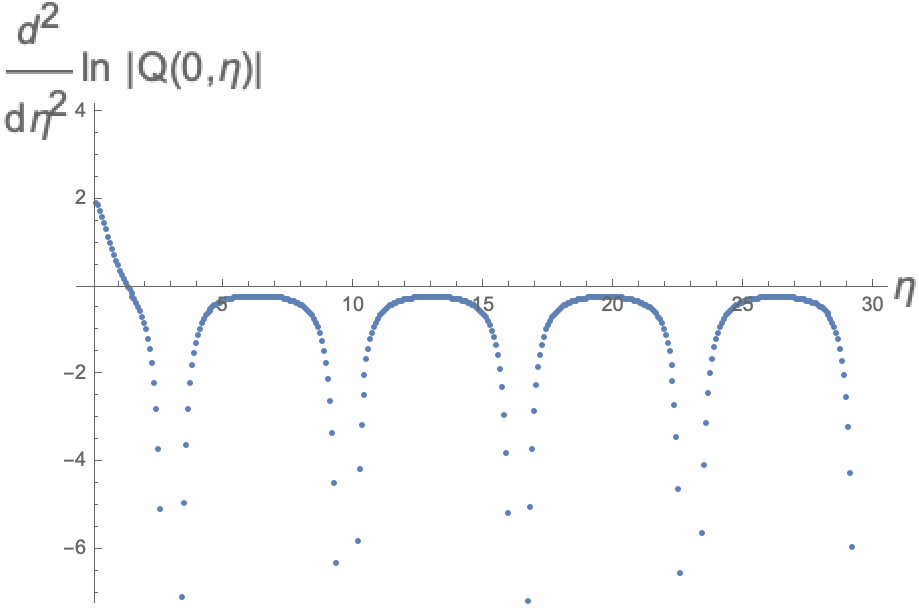}
		\caption{$\frac{d^2}{d\eta^2}\ln|Q(0,\eta)|$}
	\end{subfigure}
	\caption{Plots of $\frac{d^2}{d\eta^2}\ln|G(0,\eta)|$ and $\frac{d^2}{d\eta^2}\ln|Q(0,\eta)|$ for $N_f=3$ and $N_c =3$. Both graphs result from our numerical computation with step size $\Delta\eta = 0.075$ and $\eta_{\max} = 30$.}
\label{fig:ddlogGA}
\end{figure}

Finally, a linear regression on 
\begin{equation}
\ln\left|\frac{f(\eta)}{\cos\left(\omega \, \eta+ \varphi \right)}\right| = \alpha \, \eta + \ln K
\label{eqn:Analysis4}
\end{equation}
allows us to extract $\alpha$ from the slope of this function. For the numerical values of $G(0,\eta)$ and $Q(0,\eta)$ found in the range $\eta \in [0, \eta_{max}]$, we only use $\eta \in [0.75 \eta_{max} , \eta_{max}]$ to extract the intercepts $\alpha_G$ and $\alpha_Q$ using \eq{eqn:Analysis4}, in addition avoiding the values of $\eta$ close to the nearest cosine zero, $\eta_n = \frac{1}{\omega}\left[\frac{\pi}{2}-\varphi+\pi \, n \right]$, by at least $5\%$ of the cosine's period, $(0.05)T = \frac{\pi}{10\omega}$. This is done in order to obtain the intercept as close as possible to the asymptotic value, minimizing the errors due to numerical artifacts and oscillation.

Following the above prescription, in \fig{fig:ddlogGA} we plot $\frac{d^2}{d\eta^2}\ln\left|G(0,\eta)\right|$ and $\frac{d^2}{d\eta^2}\ln\left|Q(0,\eta)\right|$ as functions of $\eta$ for $N_f=3$. The corresponding plots for other values of $N_f$ also display the same qualitative behavior. For large $\eta$, the shapes of the graphs approach that of the function in \eq{eqn:Analysis2}, displaying periodic local maxima below the $\eta$-axis. This provides another justification for the proposed asymptotic form, \eqref{eqn:AsympNf1}, for $G(0,\eta)$ and $Q(0,\eta)$.

The method outlined above is employed to determine the asymptotic forms for $G(0,\eta)$ and $Q(0,\eta)$ at $N_f = 2,3,6$. In particular, for each $N_f$, $\Delta\eta$ and $\eta_{\max}$, we find the values of $\omega_G$, $\omega_Q$, $\varphi_G$ and $\varphi_Q$ from the largest maximum ($\eta^*$) of the graphs in \fig{fig:ddlogGA}, which, in turn, correspond to the function in \eqref{eqn:Analysis2}, in order to get as close as possible to the asymptotic behavior at large $\eta$. The frequencies are found by using \eq{max_omega}, while the phases are extracted using $\omega \, \eta^* + \varphi = \pi \, n$, with the integer $n$ adjusted as described above. The parameters $\alpha_G$ and $\alpha_Q$ can be deduced from the slope of the function in \eq{eqn:Analysis4}.

\begin{figure}[H]
	\centering
	\begin{subfigure}{.3\textwidth}
		\includegraphics[width=\textwidth]{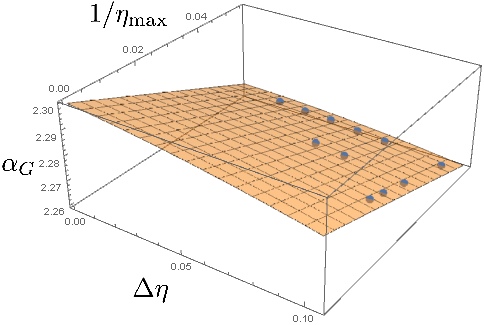}
		\caption{$\alpha_G$ for $N_f=3$}
	\end{subfigure}\;\;\;\;
	\begin{subfigure}{.3\textwidth}
		\includegraphics[width=\textwidth]{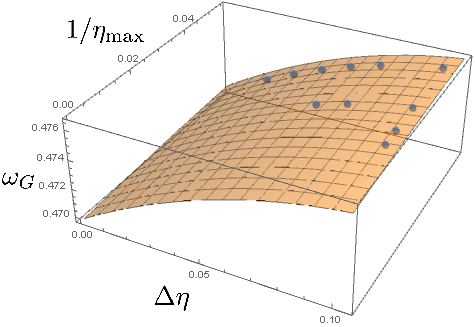}
		\caption{$\omega_G$ for $N_f=3$}
	\end{subfigure}\;\;\;\;
	\begin{subfigure}{.3\textwidth}
		\includegraphics[width=\textwidth]{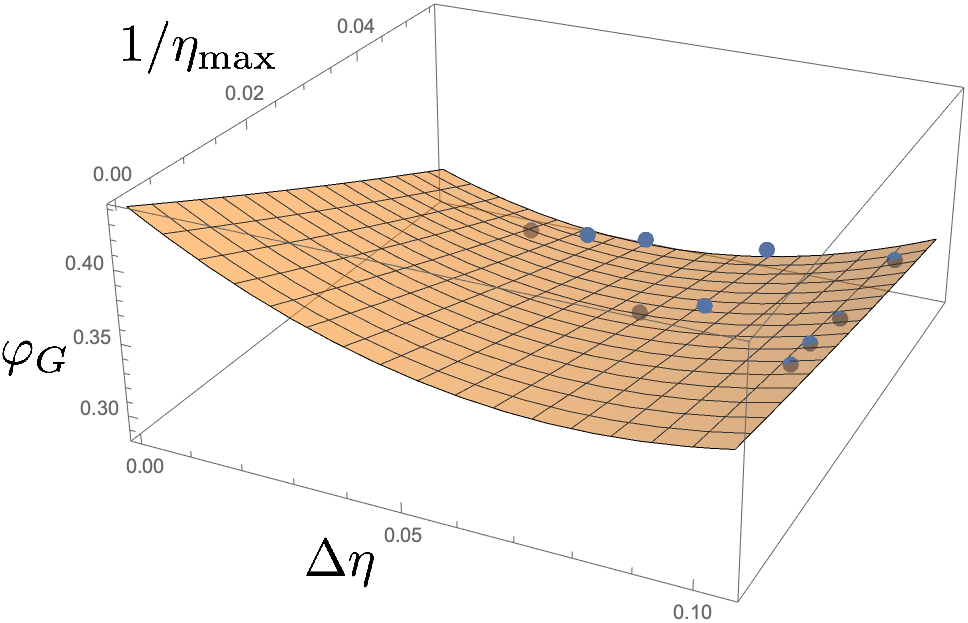}
		\caption{$\varphi_G$ for $N_f=3$}
	\end{subfigure}
	\begin{subfigure}{.3\textwidth}
		\includegraphics[width=\textwidth]{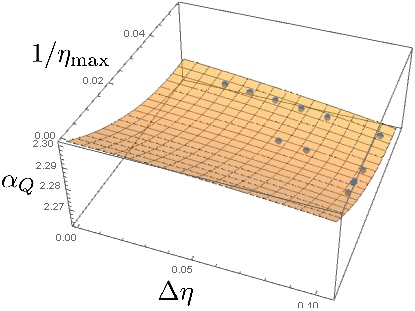}
		\caption{$\alpha_Q$ for $N_f=3$}
	\end{subfigure}\;\;\;\;
	\begin{subfigure}{.3\textwidth}
		\includegraphics[width=\textwidth]{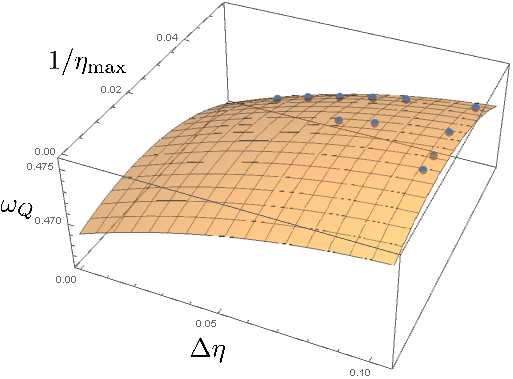}
		\caption{$\omega_Q$ for $N_f=3$}
	\end{subfigure}	\;\;\;\;
	\begin{subfigure}{.3\textwidth}
		\includegraphics[width=\textwidth]{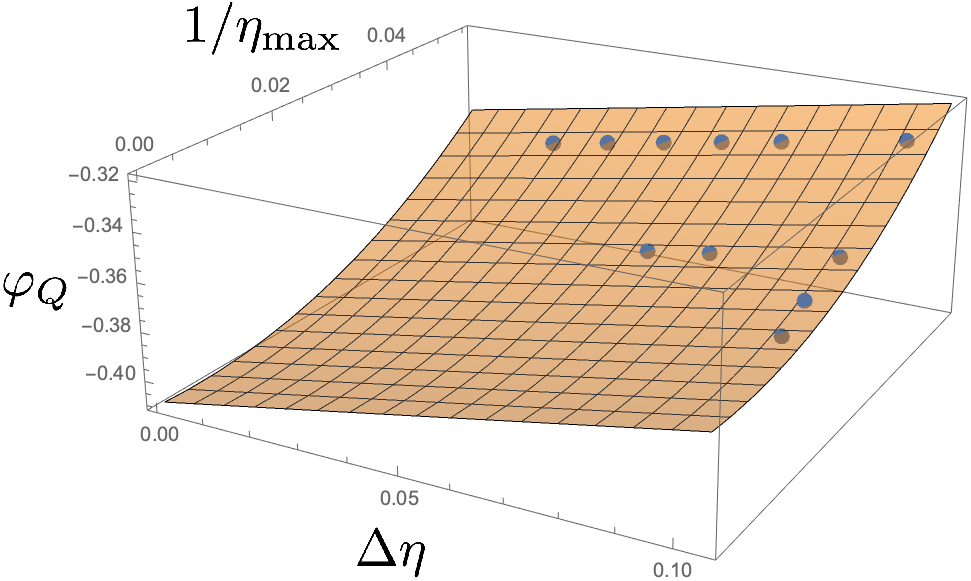}
		\caption{$\varphi_Q$ for $N_f=3$}
	\end{subfigure}
	\caption{Plots of parameters $\alpha_G$, $\alpha_Q$, $\omega_G$, $\omega_Q$, $\varphi_G$ and $\varphi_Q$ as functions of $\Delta\eta$ and $1/\eta_{\max}$ for $N_f=3$ and $N_c =3$. The dots represent our numerical evaluation, while the solid surfaces depict the best fits (quadratic in $\Delta\eta$ and $1/\eta_{\max}$) used for extrapolating to the continuum asymptotic values at $\Delta\eta=0$ and $\frac{1}{\eta_{\max}}=0$.}
\label{fig:parameters}
\end{figure}

As expected for any numerical and asymptotic solution, the resulting intercepts, $\alpha_G$ and $\alpha_Q$, frequencies, $\omega_G$ and $\omega_Q$, and initial phases, $\varphi_G$ and $\varphi_Q$, all vary slightly with the step size $\Delta\eta$ and the maximum $\eta_{\max}$ of the computation range $\eta \in [0, \eta_{max}]$. Since the exact continuum and asymptotic solution corresponds to the limit where $\Delta\eta\to 0$ and $\eta_{\max}\to +\infty$, we perform the computation with various $\Delta\eta$'s and $\eta_{\max}$'s, obtaining these parameters for each computation. Then, these values of the parameters are fitted with second-order polynomials in $\Delta\eta$ and $1/\eta_{\max}$ (cf. \cite{Kovchegov:2016weo}). For each parameter, the value of its best-fit quadratic surface at $\Delta\eta=0, \frac{1}{\eta_{\max}}=0$ is taken to be our numerical estimate for the parameter. This technique is employed to obtain the estimates of $\alpha_G$, $\alpha_A$, $\omega_G$, $\omega_A$, $\varphi_G$ and $\varphi_A$ for $N_f=2,3,6$ in the limit $\Delta\eta \to 0,  \frac{1}{\eta_{\max}} \to 0$. \fig{fig:parameters} displays by the dots the values of these parameters at $N_f=3$ for each pair of $\Delta\eta$ and $\eta_{\max}$ that we performed the computation for, together with the best-fit quadratic surfaces. For completeness, let us list the equations describing the best-fit quadratic surfaces (for $N_f=3$):
\begin{subequations}\label{eqn:Analysis5}
\begin{align}
\alpha_G\left(\Delta\eta,\eta_{\max}\right) &=  2.300 - 0.086\left(\Delta\eta\right) - 0.608\left(1/\eta_{\max}\right) - 0.472\left(\Delta\eta\right)^2 + 1.350\left(\Delta\eta/\eta_{\max}\right) - 0.108\left(1/\eta_{\max}\right)^2  , \\
\omega_G\left(\Delta\eta,\eta_{\max}\right) &= 0.470 + 0.098\left(\Delta\eta\right) + 0.018\left(1/\eta_{\max}\right) - 0.388\left(\Delta\eta\right)^2 + 0.018\left(\Delta\eta/\eta_{\max}\right) - 0.118\left(1/\eta_{\max}\right)^2 , \\
\varphi_G\left(\Delta\eta,\eta_{\max}\right) &= 0.327 + 0.516\left(\Delta\eta\right) - 1.403\left(1/\eta_{\max}\right) - 0.026\left(\Delta\eta\right)^2 - 1.618\left(\Delta\eta/\eta_{\max}\right) + 11.143\left(1/\eta_{\max}\right)^2 , \\
\alpha_Q\left(\Delta\eta,\eta_{\max}\right) &=  2.301 - 0.014\left(\Delta\eta\right) - 1.008\left(1/\eta_{\max}\right) - 0.331\left(\Delta\eta\right)^2 - 1.206\left(\Delta\eta/\eta_{\max}\right) + 9.654\left(1/\eta_{\max}\right)^2  , \\
\omega_Q\left(\Delta\eta,\eta_{\max}\right) &= 0.469 + 0.094\left(\Delta\eta\right) + 0.012\left(1/\eta_{\max}\right) - 0.385\left(\Delta\eta\right)^2 + 0.077\left(\Delta\eta/\eta_{\max}\right) - 3.052\left(1/\eta_{\max}\right)^2 , \\
\varphi_Q\left(\Delta\eta,\eta_{\max}\right) &= -0.409 + 0.386\left(\Delta\eta\right) - 0.230\left(1/\eta_{\max}\right) + 0.032\left(\Delta\eta\right)^2 - 0.795\left(\Delta\eta/\eta_{\max}\right) + 22.008\left(1/\eta_{\max}\right)^2  .
\end{align}
\end{subequations}
The qualitative features of these plots and quadratic fit functions are similar for $N_f =2, 6$, but we omit them for brevity. The resulting values of the parameters extracted with the quadratic fit are given in Table~\ref{tab:aop} of the main text.  For each of the parameters, the quadratic fit gives the value for the coefficient of determination, $R^2$, of at least $0.999$. The numerical errors shown in Table~\ref{tab:aop} are derived by taking the difference between the $\Delta\eta\to 0$ and $\eta_{\max}\to +\infty$ extrapolations of the quadratic and linear fits in $\Delta\eta$ and $1/\eta_{\max}$ to the data points in \fig{fig:parameters}.

One may wonder why $\alpha_G$ and $\alpha_Q$ in \fig{fig:parameters} approach $\Delta\eta=0$ and $\frac{1}{\eta_{\max}}=0$ from below, while in \cite{Kovchegov:2016weo} the intercept approached the same limit from above. To better understand the difference, we re-ran the large-$N_c$ evolution simulations done in \cite{Kovchegov:2016weo} with the initial conditions different from those used in \cite{Kovchegov:2016weo}. We used $G^{(0)} =1$, while in \cite{Kovchegov:2016weo} Born-level initial conditions \eqref{BornLvIC} were employed. Using the trivial initial condition $G^{(0)} =1$ resulted in the intercept approaching the $\Delta\eta=0$ and $\frac{1}{\eta_{\max}}=0$ limit from below for the large-$N_c$ evolution. We thus conclude that, while the asymptotic and continuum value of the intercept appears to be independent of the initial conditions, the approach to this intercept in finite-step-size and finite-$\eta$-range numerical simulations appears to depend on the initial conditions.

\begin{figure}[H]
	\centering
	\begin{subfigure}{.3\textwidth}
		\includegraphics[width=\textwidth]{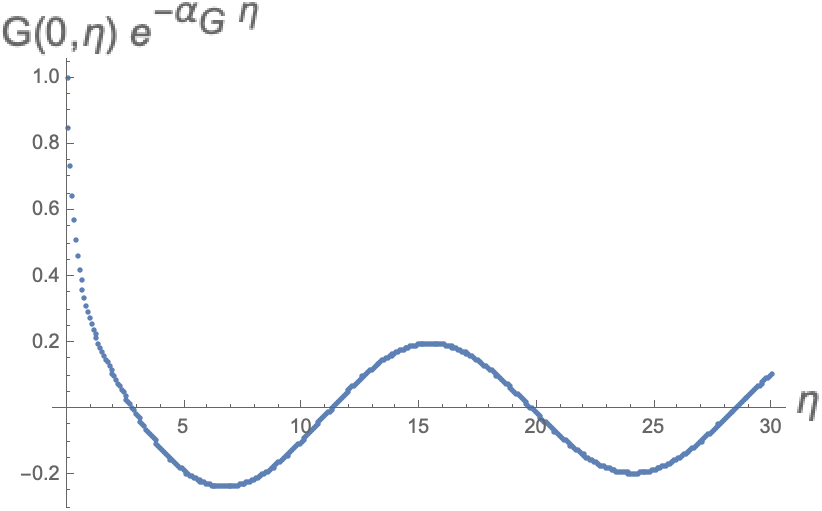}
		\caption{$e^{-\alpha_G\eta}G(0,\eta)$ at $N_f=2$}
	\end{subfigure}\;~\;~\;\;
	\begin{subfigure}{.3\textwidth}
		\includegraphics[width=\textwidth]{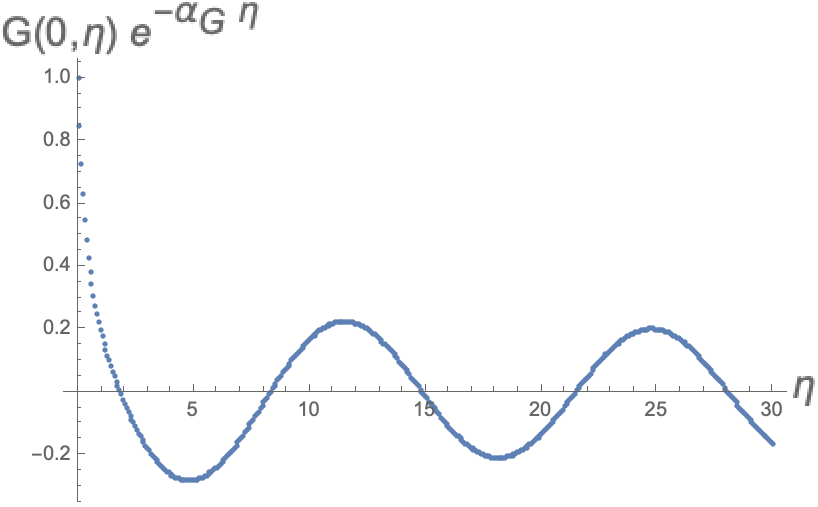}
		\caption{$e^{-\alpha_G\eta}G(0,\eta)$ at $N_f=3$}
	\end{subfigure}\;~\;~\;\;
	\begin{subfigure}{.3\textwidth}
		\includegraphics[width=\textwidth]{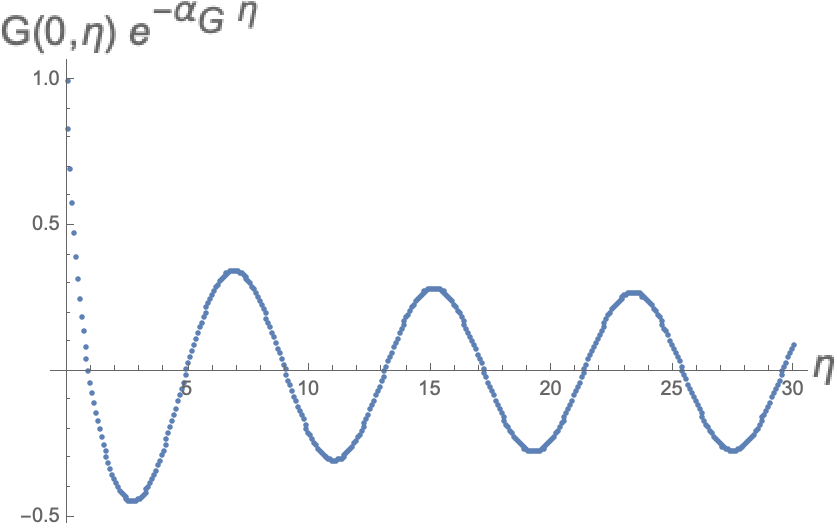}
		\caption{$e^{-\alpha_G\eta}G(0,\eta)$ at $N_f=6$}
	\end{subfigure}\;~\;~\;\;
	\begin{subfigure}{.3\textwidth}
		\includegraphics[width=\textwidth]{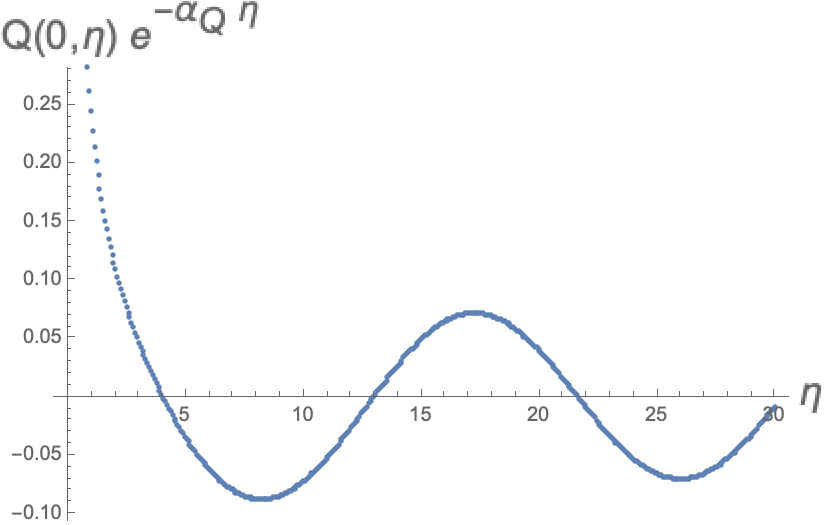}
		\caption{$e^{-\alpha_Q\eta}Q(0,\eta)$ at $N_f=2$}
	\end{subfigure}\;~\;~\;\;
	\begin{subfigure}{.3\textwidth}
		\includegraphics[width=\textwidth]{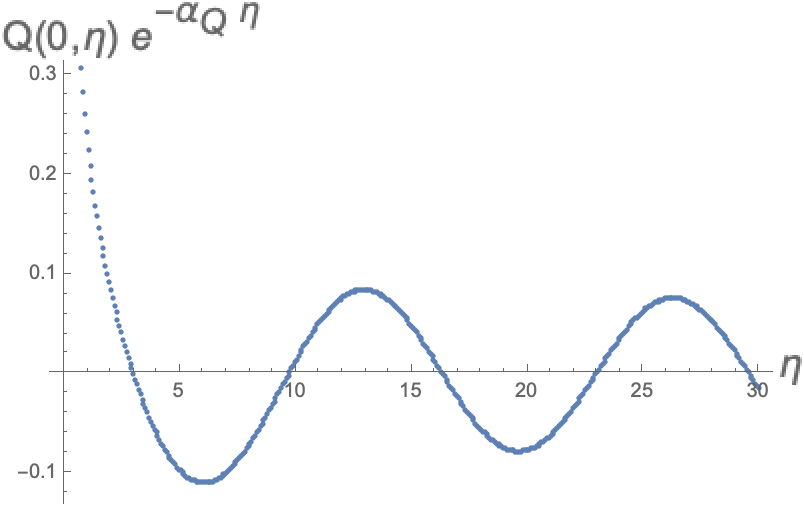}
		\caption{$e^{-\alpha_Q\eta}Q(0,\eta)$ at $N_f=3$}
	\end{subfigure}\;~\;~\;\;
	\begin{subfigure}{.3\textwidth}
		\includegraphics[width=\textwidth]{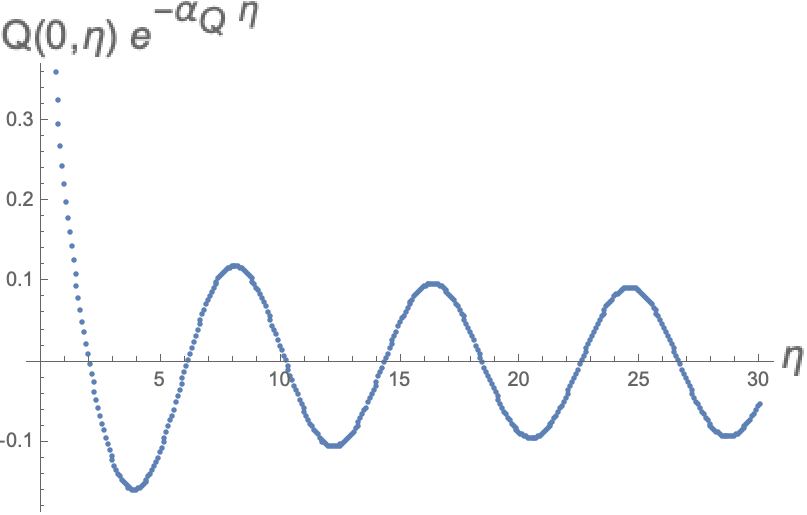}
		\caption{$e^{-\alpha_Q\eta}Q(0,\eta)$ at $N_f=6$}
	\end{subfigure}
	\caption{Plots of $e^{-\alpha_G\eta}G(0,\eta)$ and $e^{-\alpha_Q\eta}Q(0,\eta)$ at $N_f=2,3,6$ and $N_c =3$. All the graphs are numerically computed with step size $\Delta\eta = 0.075$ and $\eta_{\max} = 30$.}
\label{fig:oscillating}
\end{figure}

As a final cross check for the asymptotic form \eqref{eqn:AsympNf1}, we plot $e^{-\alpha_G\eta}G(0,\eta)$ and $e^{-\alpha_Q\eta}Q(0,\eta)$ versus $\eta$ in \fig{fig:oscillating}. We see that the functions display clear sinusoidal pattern for $\eta \gtrsim 10$, demonstrating sinusoidal oscillation in the large-$\eta$ asymptotics, as expected from the ansatz \eqref{eqn:AsympNf1}.

\section{Analysis of the numerical results for $\Delta \Sigma$}

\label{sec:analysis_DS}

\begin{figure}[h]
	\centering
	\begin{subfigure}{.3\textwidth}
		\includegraphics[width=\textwidth]{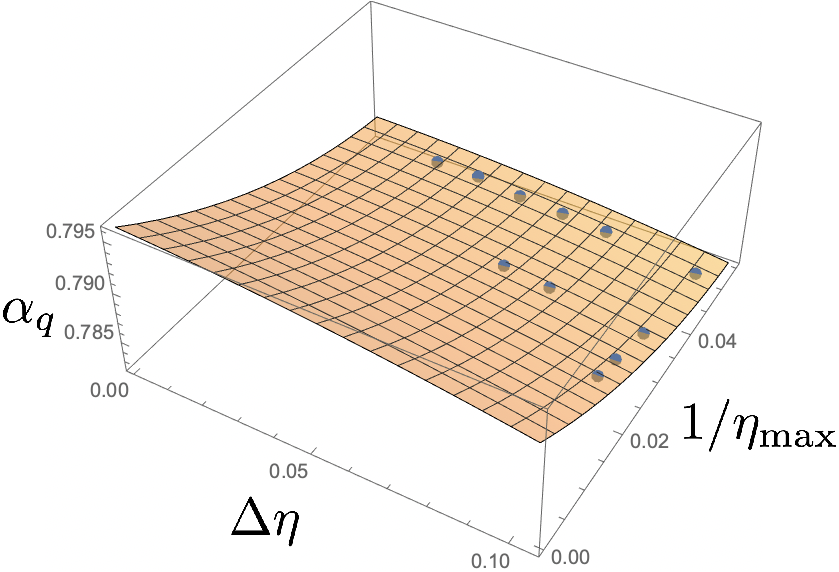}
		\caption{$\alpha_{q}$ for $N_f=3$}
	\end{subfigure}\;\;\;\;
	\begin{subfigure}{.3\textwidth}
		\includegraphics[width=\textwidth]{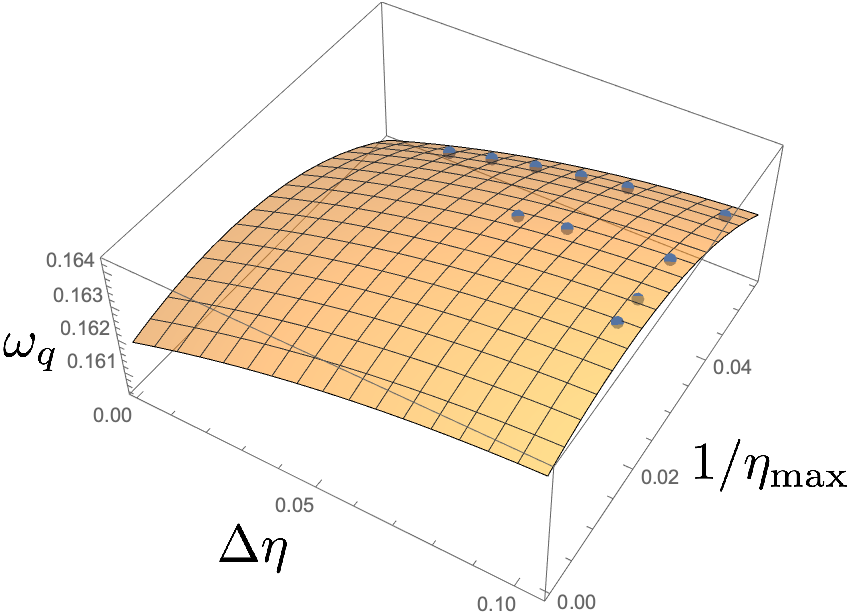}
		\caption{$\omega_{q}$ for $N_f=3$}
	\end{subfigure}\;\;\;\;
	\begin{subfigure}{.3\textwidth}
		\includegraphics[width=\textwidth]{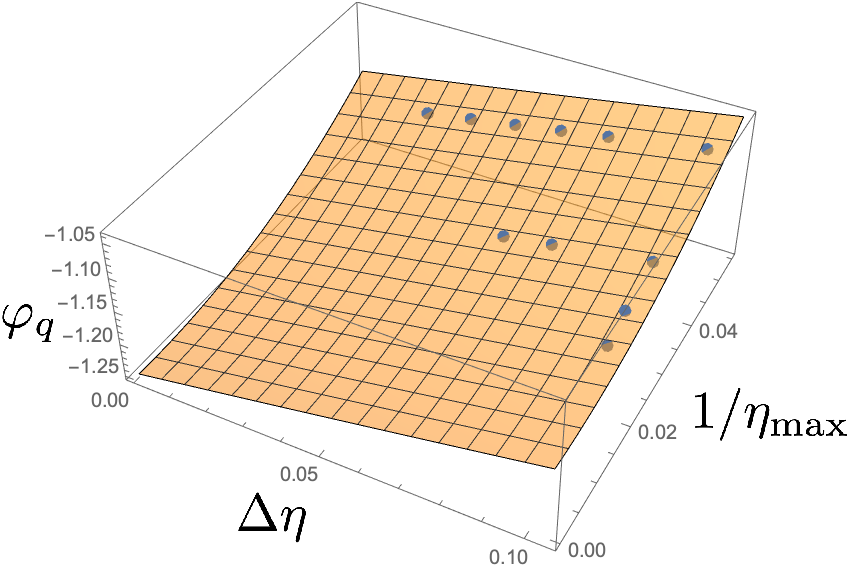}
		\caption{$\varphi_{q}$ for $N_f=3$}
	\end{subfigure}
	\caption{Plots of parameters $\alpha_{q}$, $\omega_{q}$ and $\varphi_{q}$ as functions of $\Delta\eta$ and $1/\eta_{\max}$ for $N_f=3$ and $N_c =3$. The dots represent our numerical evaluation, while the solid surfaces depict the best fits (quadratic in $\Delta\eta$ and $1/\eta_{\max}$) used for extrapolating to the continuum asymptotic values at $\Delta\eta=0$ and $\frac{1}{\eta_{\max}}=0$.}
\label{fig:parameters2}
\end{figure}

Here we use the fitting method outlined in Appendix~\ref{sec:analysis} to compute the parameters given in \eq{eqn:DelSigmaAsymp2} describing the small-$x$ asymptotics of $\Delta\Sigma (x, Q^2)$ in \eq{eqn:DelSigmaAsymp}. The only difference is that the variable $\eta$ in Eqs.~\eqref{eqn:Analysis1}-\eqref{eqn:Analysis4} now becomes $\sqrt{\frac{\as\;N_c}{2\pi}}\;\ln\frac{1}{x}$. At the end, the parameters $\alpha_q$, $\omega_q$, and $\varphi_q$ are extracted by using the following quadratic best-fit surfaces for $\as = 0.25$, resulting in the values listed in \eq{eqn:DelSigmaAsymp2}:
\begin{subequations}\label{eqn:Analysis6}
\begin{align}
\alpha_{q}\left(\Delta\eta,\eta_{\max}\right) &=  0.796 - 0.016\left(\Delta\eta\right) - 0.362\left(1/\eta_{\max}\right) - 0.192\left(\Delta\eta\right)^2 + 0.216\left(\Delta\eta/\eta_{\max}\right) + 2.852\left(1/\eta_{\max}\right)^2 ,  \\
\omega_{q}\left(\Delta\eta,\eta_{\max}\right) &= 0.162 + 0.032\left(\Delta\eta\right) + 0.052\left(1/\eta_{\max}\right) - 0.132\left(\Delta\eta\right)^2 + 0.024\left(\Delta\eta/\eta_{\max}\right) - 1.444\left(1/\eta_{\max}\right)^2 , \\
\varphi_{q}\left(\Delta\eta,\eta_{\max}\right) &= -1.25 + 0.96\left(\Delta\eta\right) - 0.36\left(1/\eta_{\max}\right) + 0.32\left(\Delta\eta\right)^2 - 1.66\left(\Delta\eta/\eta_{\max}\right) + 45.43\left(1/\eta_{\max}\right)^2 .
\end{align}
\end{subequations}
Similarly, for all the parameters, we obtain the coefficient of determination $R^2$ of at least $0.996$. \fig{fig:parameters2} displays by the dots the values of these parameters for each pair of $\Delta\eta$ and $\eta_{\max}$ that we performed the computation for, together with the best-fit quadratic surfaces. The error bars in \eq{eqn:DelSigmaAsymp2} are also calculated by the difference between the linear and quadratic extrapolations to $\Delta\eta=0$ and $\frac{1}{\eta_{\max}}=0$.



\providecommand{\href}[2]{#2}\begingroup\raggedright\endgroup

\end{document}